\documentclass[fleqn,usenatbib]{mnras}

\usepackage[T1]{fontenc}

\DeclareRobustCommand{\VAN}[3]{#2}
\let\VANthebibliography\thebibliography
\def\thebibliography{\DeclareRobustCommand{\VAN}[3]{##3}\VANthebibliography}

\usepackage{graphicx}	
\usepackage{amsmath}	
\usepackage{amssymb}	
\usepackage{soul}
\usepackage{newtxtext,newtxmath}

\title[HNCO in the hot molecular core G331]{Isocyanic acid (HNCO) in the Hot Molecular Core G331.512-0.103: Observations and Chemical Modelling}

\author[Carla M. Canelo et al.]{
Carla M. Canelo,$^{1}$\thanks{E-mail: camcanelo@gmail.com}
Leonardo Bronfman,$^{2}$
Edgar Mendoza,$^{1,3}$
Nicolas Duronea,$^{4}$
Manuel Merello,$^{2}$
\newauthor Miguel Carvajal, $^{5,6}$
Am\^ancio C.S. Fria\c{c}a,$^{1}$
and Jacques Lepine$^{1}$
\\
$^{1}$Departamento de Astronomia, Instituto de Astronomia, Geof\'isica e Ci\^encias Atmosf\'ericas, Universidade de S\~ao Paulo, S\~ao Paulo, Brazil\\
$^{2}$Departamento de Astronom{\'{\i}}a, Universidad de Chile, Casilla 36-D, Santiago de Chile, Chile \\
$^{3}$Observat\'orio do Valongo, Universidade Federal do Rio de Janeiro, Ladeira Pedro Ant\^onio, 43, Rio de Janeiro – RJ 20.080-090, Brazil\\
$^{4}$Instituto de Astrof\'isica de La Plata (UNLP - CONICET), La Plata, Argentina\\
$^{5}$Dept. Ciencias Integradas, Facultad de Ciencias Experimentales, Centro de Estudios Avanzados en F\'isica, Matem\'atica y Computaci\'on,\\ Unidad Asociada GIFMAN, CSIC-UHU, Universidad de Huelva, Spain \\
$^{6}$Instituto Universitario Carlos I de F\'{\i}sica Te\'orica y Computacional, Universidad de Granada, Spain
}

\date{Accepted 2021 April 20. Received 2021 April 20; in original form 2021 January 27.}

\pubyear{2020}

\begin{document}
\label{firstpage}
\pagerange{\pageref{firstpage}--\pageref{lastpage}}
\maketitle

\begin{abstract}
Isocyanic acid (HNCO) is a simple molecule with a potential to form prebiotic and complex organic species. Using a spectral survey collected with the Atacama Pathfinder EXperiment (APEX), in this work we report the detection of 42 transitions of HNCO in the hot molecular core/outflow G331.512-0.103 (hereafter G331). The spectral lines were observed in the frequency interval $\sim$~160--355 GHz. By means of Local Thermodynamic Equilibrium (LTE) analyses, applying the rotational diagram method, we studied the excitation conditions of HNCO. The excitation temperature and column density are estimated to be $T_{ex}$= 58.8 $\pm$ 2.7~K and $N$ = (3.7 $\pm$ 0.5) $\times$ 10$^{15}$~cm$^{-2}$, considering beam dilution effects.  The derived relative abundance is between (3.8 $\pm$ 0.5) $\times $10$^{-9}$ and (1.4 $\pm$ 0.2) $\times $10$^{-8}$. In comparison with other hot molecular cores, our column densities and abundances are in agreement. An update of the internal partition functions of the four CHNO isomers: HNCO; cyanic acid, HOCN; fulminic acid, HCNO; and isofulminic acid, HONC is provided. We also used the astrochemical code \textsc{Nautilus} to model and discuss HNCO abundances. The simulations could reproduce the abundances with a simple zero-dimensional model at a temperature of 60~K and for a chemical age of $\sim$10$^5$~years, which is larger than the estimated dynamical age for G331. This result could suggest the need for a more robust model and even the revision of chemical reactions associated with HNCO.

\end{abstract}

\begin{keywords}
ISM: molecules -- astrochemistry -- methods: observational -- molecular data -- molecular processes
\end{keywords}



\section{Introduction}
\label{intro:G331}

The isocyanic acid (HNCO) is the simplest molecule containing four of the six biogenic elements: carbon, hydrogen, oxygen, nitrogen, phosphorus and sulphur (abbreviated as CHONPS) that are present in all living beings.  Although it has only four atoms, HNCO could be precursor of other prebiotic and complex organic molecules (COMs\footnote{As discussed in \citet{Herbst2009}, the term complex is labelled by astronomers if not by chemist, and makes a reference to interstellar molecules with 6 or more atoms including the element carbon.}), which   are of great astrochemical and astrobiological interest due to their potential to form molecules such as amino acids, sugars and nucleobases \citep[e.g.][and references therein]{Gorai2020}. HNCO molecules are included as a peptide bond [–(H)N–C(O)–] between any two single amino acid, such as glycine \citep{Fedoseev15}. Moreover, recent experiments,  simulating interstellar ice chemical processes, have shown that HNCO and CH$_4$ can form  molecules with peptide bonds in solid state under far-UV irradiation \citep{Ligterink18}. The observation of COMs in astrophysical environments, along with the study of their formation and destruction pathways, are important for a better comprehension of the origins of life on Earth and elsewhere.

The potential interest of HNCO in radio astronomy has generated and stimulated debates, specially about its formation and detection. Among the first experimental works, \citet{Jones1950a} and \citet{Jones1950b} made experimental investigations of the infrared and microwave spectra of HNCO and HNCS. \citet{Hocking1972} and \citet{Hocking1975} carried out laboratory measurements to study the HNCO microwave and millimeter spectrum. Concurrently, \citet{snyder72} reported the results of a survey for HNCO in Galactic molecular sources, concluding about its detection in Sgr~B2. HNCO is a near prolate asymmetric top molecule with a linearity barrier at 1899~cm$^{-1}$~\citep{Niedenhoff1995}; its rotational levels undergo a hyperfine splitting caused by the nuclear spin of nitrogen~\citep{Lapinov2007}. The rotational levels of HNCO  are assigned by the spectroscopic notation for the asymmetric top, such as $J_{K_a,K_c}$, where $J$ is the rotational angular momentum, and $K_a$ and $K_c$ are the projections of $J$ to the molecule fixed principal $a$ and $c$-axes, respectively   \citep{Hocking1974,Niedenhoff1995,Zinchenko2000}. 

\citet{Churchwell1986} found that the most likely excitation mechanism of HNCO in Sgr~B2 might be radiative rather than collisional, so the molecule can be an optimal tracer of the far infrared field (at $\lambda \thickapprox$~110 and 330~$\mu$m). As a consequence of the energy structure and short lifetimes of the excited $K_a$ ladders of HNCO, its populational diagram can be substantially induced by far-infrared radiation. In a study based on  observations of HNCO (5$_{0,5}$--4$_{0,4}$) in massive star forming regions, \citet{Li2013} found that the HNCO emission show signs of embedded mid-infrared or far-infrared sources, and also a correlation with emission of HC$_3$N.

As we mentioned previously, the first detection of HNCO in the interstellar medium (ISM) was towards the molecular cloud complex Sgr B2 \citep{snyder72}. Since then, HNCO has been periodically reported in several astronomical objects, from comets up to external galaxies. Based on sub-mm observations of comets, \citet{Biver06} carried out a comparative study of molecules such as CH$_3$OH, H$_2$CO, HNCO, among others, towards the comets C/1999 T1 (McNaught-Hartley), C/2001 A2 (LINEAR), C/2000 WM1 (LINEAR) and 153P/Ikeya-Zhang. \citet{Lis1997} performed observations of Comet C/1996 B2 (Hyakutake) and reported for the first time the HNCO 16$_{0,16}$--15$_{0,15}$ transition, at $\nu \thickapprox$~351633.257~MHz, in a comet. \citet{Crovisier98} discussed the presence of HNCO in comets Hyakutake and Hale-Bopp; its synthesis could be associated with the OCN$^-$ anion. In the context of planetary atmospheres, based on  Herschel/HIFI-PACS and Cassini/CIRS studies of Titan's atmosphere, \citet{Dob14} carried out photochemical models to predict the presence of compounds such as NO, HNCO and N$_2$O. In low-mass protostellar objects, HNCO was studied along with NH$_2$CHO in solar-type protostars and protostellar shocks \citep{Mendoza14,Lopez2015,Lopez2019}. 

More recently, \citet{Hernandez19} observed HNCO towards IRAS 16293-2422 and computed models using the astrochemical code \textsc{Nautilus}. Their study included observations from several telescopes including the Atacama Pathfinder Experiment (APEX). Considering evolved objects, HNCO was observed in the oxygen-rich circumstellar envelope around an intermediate-mass evolved star, OH231.8+4.2, which harbors a bipolar molecular outflow  \citep{Prieto15}. In the context of high mass star forming regions, \citet{Zinchenko2000} surveyed 81 dense high-mass star-forming regions and emission lines of HNCO were detected towards 57 sources, with a tentative detection of HN$^{13}$CO (10$_{0,10}$--9$_{0,9}$) at $\sim$~220~GHz in G310.12-0.20. \citet{Gorai2020} carried out a study of HNCO, NH$_2$CHO and CH$_3$NCO in the hot molecular core G10.47+0.03, finding that they are chemically linked with each other according to their data analysis results and chemical models. In the same source, \citet{Wyrowski1999} had previously reported the first detection of vibrationally excited HNCO, as part of a study that was mainly focused on HC$_3$N. In Sgr~B2 and Orion~KL, \citet{Turner91} analysed several transitions of HNCO in the frequency interval $\sim$~87--220~GHz. In extragalactic sources, \citet{Nguyen1991} reported the first detection of HNCO in external galaxies. They observed the 4$_{0,4}$--3$_{0,3}$ and 6$_{0,6}$--5$_{0,5}$ transitions towards NGC~253 using the IRAM-30m observatory. \citet{Meier05} observed emission of HNCO towards the nearby spiral galaxy IC 342, finding a correlation with another tracer as CH$_3$OH. \citet{Martin06} analysed the physical conditions of HNCO, among other chemical species, using a 2~mm spectral survey towards NGC~253. \citet{Martin09} determined HNCO abundances in galaxies, concluding that HNCO is a good tracer to diagnose the evolutionary state of nuclear starburst. 

G331 has proven to be very rich source in molecular emission \citep[e.g.][]{Bronfman08,Merello13a,Mendoza18,Duronea19,Hervias19} and the analysis of HNCO could be an important tool not only to understand the physical properties of the source but also to complement the inventory of Galactic sources where this molecule has been detected. Moreover, the present study could increase the chemical knowledge of the molecular species and their reaction networks present in G331. 

In this work, we focus on the identification of HNCO in the hot molecular core G331, which is  one of the most luminous and energetic outflows known in the Galaxy. It is located towards the brightest and most massive dust condensation of the star-forming region of G331.5-0.1, placed in the Norma spiral arm at a heliocentric distance of approximately 7.5~kpc \citep{Merello13b}. The mass of the core is about 40~$M_{\sun}$ while the mass of each outflow lobe is about 25~$M_{\sun}$ \citep{Hervias19}. G331 also presents a momentum of $\sim$ 2.4$\times$10$^3$~$M_{\sun}$~km~s$^{-1}$ and a kinetic energy of  $\sim $1.4$\times$10$^{48}$~erg \citep{Bronfman08}. Such values are expected in flows driven by young massive stellar objects with L$_{bol} \sim$ 1$\times$10$^5$~$L_{\sun}$  \citep[e.g.][]{Bally2016}.  Observations in the ALMA band 7 receiver revealed a main hydrogen density of the core of $\sim$ 5~$\times$~10$^6$~cm$^{-3}$ and a temperature around 70~K \citep{Hervias19}. Moreover, the observed SiO and SO$_2$ molecular lines also traced a denser region with temperatures up to 200~K which suggests an expanding shell-like structure. 

Another approach to understand the chemical network of HNCO in G331 is to simulate its abundance with an astrochemical model. The evidence of ammonium salts in comet 67P/Churyumov-Gerasimenko can suggest that species like halides and HNCO are in the form of salt, presenting high sublimation temperatures which would not allow their detection in the gas phase and could explain the relatively low interstellar abundances of HNCO \citep{Altwegg20}.  In this sense, the \textsc{Nautilus} code \citep{nautilus} considers both gas and grain reactions to perform time-dependent simulations of molecular abundances in hot and cold molecular cores, which can allow an overview of the HNCO chemistry in G331. 

The article is divided as following: Section~\ref{sec:methods} presents the methodology of the observations and data analysis; Section~\ref{sec:results} exhibits the detected lines and their physical analysis; these results are also discussed in Section~\ref{sec:disc} together with the \textsc{Nautilus} simulations; and finally,  the summary and conclusions of this work are presented in Section~\ref{sec:conc}.

\section{Methodology}
\label{sec:methods}

The observations have been obtained with the APEX telescope\footnote{This publication is based on data acquired with the Atacama Pathfinder Experiment (APEX), over various semesters between 2014 and 2019, under projects IDs  C-094.F-9709B-2014, C-097.F-9710A-2016, C-099.F-9702A-2017, C-0102.F-9702B-2018 and C-0104.F-9703B-2019. APEX is a collaboration between the Max-Planck-Institut fur Radioastronomie, the European Southern Observatory, and the Onsala Space Observatory.} \citep{gu06} using the single point mode towards the coordinates RA:DEC = 16$^h$12$^m$10.1$^s$, $-$51$^{\circ}$28$^{\prime}$38.1$^{\prime\prime}$.  The spectral line setups were collected over various semesters between 2014 and 2019. We have used APEX-1 and APEX-2 receivers of the  Swedish Heterodyne Facility Instrument (SHeFI; \citealt{vas08}) to observe bands within the intervals 213--275~GHz and 267--378~GHz, respectively. As backend, the eXtended bandwidth Fast Fourier Transform Spectrometer2 (XFFTS2) has been used. The spectral resolution, corresponding to a velocity resolution, has been  obtained between $\sim$~0.15 and 0.25~km~s$^{-1}$ for a noise level of $\sim$~30 mK. This paper includes results from the last observations conducted with the SEPIA B5 receiver \citep{belitsky2018}, since various lines of HNCO were discovered within the frequency interval covered by that receiver (159-211~GHz). For further clarity, the resolution of the spectra exhibited in the present work has been  degraded to a common value of 1~km~s$^{-1}$. The original intensity, obtained in scale of antenna temperature ($T_a$), has been converted to the main-beam temperature ($T_{mb}$) scale using main-beam efficiency values of $\eta_{mb} \thickapprox$ 0.80, 0.75 and 0.73 at the frequencies 208, 230 and 352~GHz~\footnote{\url{http://www.apex-telescope.org/telescope/efficiency/}}, respectively, whose Half Power Beam Width (HPBW) values are within $\sim$~17--30~arcsec. 

The data reduction and line identification have been carried out using the CLASS/GILDAS\footnote{\url{https://www.iram.fr/IRAMFR/GILDAS/}} and CASSIS\footnote{\url{http://cassis.irap.omp.eu/}} softwares. Lines have been analysed using Spectroscopic databases such as the NIST\footnote{\url{http://physics.nist.gov/cgi-bin/micro/table5/start.pl}}  Recommended Rest Frequencies for Observed Interstellar Molecular Microwave Transitions \citep{Lovas2004},  CDMS\footnote{\url{https://www.astro.uni-koeln.de/cdms/catalog}} \citep{end2016}, JPL\footnote{\url{https://spec.jpl.nasa.gov/}} \citep{pic1998} and Splatalogue\footnote{\url{https://www.cv.nrao.edu/php/splat/}}. The radiative analyses have been carried out assuming Local Thermodynamic Equilibrium (LTE), excitation temperatures and column densities were estimated from rotational diagrams \citep{gol1999}. The rotational diagrams were constructed using CASSIS, which requires  a given calibration uncertainty to compute rotational temperatures and column densities with their respective errors. A calibration uncertainty of 30~per~cent \citep{dum2010} was considered in this work. Descriptions on the methodology and data analysis have been presented in previous works (e.g. \citealt{Mendoza18,Duronea19}).


\section{Results}
\label{sec:results}

In general, the spectral lines of HNCO have been identified over the 3$\sigma$ level. Lines with ladders $K_a$~=~0, 1 and 2 have been detected and independently analysed in Sec.~\ref{subsec:qual-analysis}, the spectra are exhibited in Fig.~\ref{fig:k0}, \ref{fig:k1} and \ref{fig:k2.1}. Across the different $K_a$-ladder lines of HNCO, it is noted that some of them are more sensitive to the presence of the outflow. In order to illustrate that, and considering the case of HNCO lines affected by an outflow in the  circumstellar envelope of the late star OH 231.8+4.2 \citep{Prieto15}, we examine in detail two HNCO lines in G331 whose spectral profiles are displayed in Fig.~\ref{fig:i4}. 

We obtained Gaussian fit parameters for the identified HNCO lines. The results are presented in Table~\ref{tab:lines-v2}, where we list both the spectroscopic and fit parameters with their respective uncertainties. Concerning the HNCO line intensities and integrated areas, their values decrease inversely with the $K_a$ ladder number, as it was also discussed in previous observational studies \citep{Churchwell1986,Zinchenko2000,Prieto15}. While the HNCO $K_a$~=~0 lines are  strong enough to be detected, those with $K_a$~=~2 require  a more carefully analysis due to the weakness of the emission. We have performed the line identification adopting three standard criteria (e.g. \citealt{Snyder2005}), namely: {\it a)} the agreement between the rest frequency of an assigned transition with the frequency at the local source rest velocity of the source, which for G331 is $V_{lsr}\simeq$~-90~km~s$^{-1}$; {\it b)} lines with signal to noise ratio over the limit of detection (>$3\sigma$); {\it c)} considering the relative intensities of predicted lines, e.g. under LTE conditions, and comparing with previous observational studies.

\subsection{Qualitative analysis}
\label{subsec:qual-analysis}

\subsubsection*{HNCO $K_a$~=~0}

\begin{table*}

\caption{Lines of HNCO detected in G331. The transitions are presented following the order of the $K_a$-ladder numbers, $K_a$~=~0, $K_a$~=~1 and $K_a$~=~2. The integrated flux (K km~s$^{-1}$), line width (FWHM, km~s$^{-1}$) and line position ($V_{lsr}$, km~s$^{-1}$) are  estimated from Gaussian fits.}

\label{tab:lines-v2}

\centering
\begin{tabular}{cccc|ccccl}
\hline
Frequency   & Transition    & $A_{ul}$      	& $E_u$ 	& Integrated Flux   		& FWHM            		& $V_{lsr}$ 	& NIST\\
(MHz)   & $(J_{k_a,k_c})_u-(J_{k_a,k_c})_l$  	&  (10$^{-5}$   s$^{-1}$)   	& (K)      	& (K km s$^{-1}$)   		&  (km s$^{-1}$) 	& (km s$^{-1}$) & reference \\
\hline
175843.695	&	8$_{0,8}$ -- 7$_{0,7}$ 	&	7.43	&	37.98 	&	3.71	$\pm$	0.03	&	5.91	 $\pm$ 	0.05	&	-90.23	$\pm$	0.02 & 	--- \\
197821.461	&	9 $_{0,9}$ -- 8$_{0,8}$	&	10.70	&	47.47	&	3.63	$\pm$	0.03	&	5.86 $\pm$	0.07	&	-90.36  $\pm$	0.03 & 	--- \\
219798.274 & 10$_{0,10}$ -- 9$_{0,9}$ & 14.70 & 58.02 & 4.53 $\pm$ 0.05 & 5.53 $\pm$ 0.07 & -90.50 $\pm$ 0.03 & [1] \\
241774.032 & 11$_{0,11}$ -- 10$_{0,10}$ & 19.60 & 69.62 & 5.35 $\pm$ 0.33 & 5.71 $\pm$ 0.44 & -90.47 $\pm$ 0.17 &  [3] \\
263748.625 & 12$_{0,12}$ -- 11$_{0,11}$ & 25.60 & 82.28 & 4.86 $\pm$ 0.08 & 5.95 $\pm$ 0.12 & -90.66 $\pm$ 0.04  & [1] \\
285721.951	&	13$_{0,13}$ -- 12$_{0,12}$ 	&	32.60	&	95.99 	&	7.90 $\pm$ 0.10	&	6.70 $\pm$ 0.10	&	-90.76	$\pm$	0.04 & ---	\\
307693.905	&	14$_{0,14}$ -- 13$_{0,13}$ 	&	40.90  	&	110.76	&	7.24	$\pm$	0.03	&	6.07	$\pm$	0.03	&	-90.56	$\pm$	0.01 & ---	\\
329664.367 & 15$_{0,15}$ -- 14$_{0,14}$ & 50.40 & 126.58 & 8.02 $\pm$ 0.07 & 6.42 $\pm$ 0.08 & -90.64 $\pm$ 0.03 & --- \\
351633.257  & 16$_{0,16}$ -- 15$_{0,15}$ & 61.30 & 143.46 & 6.79 $\pm$ 0.06 & 6.60 $\pm$ 0.08 & -90.51 $\pm$ 0.03 & [7] \\
\hline
175189.027	&	8$_{1,8}$ -- 7$_{1,7}$ 	&	7.16	&	81.11 	&	1.01	$\pm$	0.01	&	5.08	$\pm$	0.07	&	-90.93	$\pm$	0.03 & 	--- \\
176472.191	&	8$_{1,7}$ -- 7$_{1,6}$ 	&	7.31	&	81.39	&	0.76	$\pm$	0.01	&	4.60	$\pm$	0.10	&	-90.97	$\pm$	0.03 & 	--- \\
197085.416	&	9$_{1,9}$ -- 8$_{1,8}$ 	&	10.30	&	90.57	&	0.68	$\pm$	0.01	&	4.42	$\pm$	0.08	&	-90.99	$\pm$	0.03 & 	--- \\
198528.881	&	   9$_{1,8}$ -- 8$_{1,7}$ 	&	10.50	&	90.92 	&	1.01	$\pm$	0.01	&	5.60 $\pm$ 0.10	&	-90.85	$\pm$	0.04 & 	--- \\
218981.009 & 10$_{1,10}$ -- 9$_{1,9}$ & 14.20 & 101.08 & 1.75 $\pm$ 0.05 & 5.73 $\pm$ 0.21 & -90.67 $\pm$ 0.07 &  [1] \\
220584.751 & 10$_{1,9}$ -- 9$_{1,8}$ & 14.50 & 101.50 & 1.73 $\pm$ 0.06 & 6.29 $\pm$ 0.31 & -90.79 $\pm$ 0.10 & [2] \\
240875.727 & 11$_{1,11}$ -- 10$_{1,10}$ & 19.00 & 112.64 & 1.56 $\pm$ 0.03 & 5.48 $\pm$ 0.12 & -90.91 $\pm$ 0.05 & [3] \\
242639.704 & 11$_{1,10}$ -- 10$_{1,9}$ & 19.50 & 113.15 & 1.74 $\pm$ 0.04 & 5.52 $\pm$ 0.13 & -90.67 $\pm$ 0.05 & [3] \\
262769.477 & 12$_{1,12}$ -- 11$_{1,11}$ & 24.80 & 125.25 &  1.80 $\pm$ 0.10 & 5.80 $\pm$ 0.50 & -90.80  $\pm$  0.20 & --  \\
264693.655 & 12$_{1,11}$ -- 11$_{1,10}$ & 25.40 & 125.85 &  1.59 $\pm$ 0.04 & 5.21 $\pm$ 0.19 & -90.85 $\pm$ 0.07 & [4] \\
284662.172	&	13$_{1,13}$ -- 12$_{1,12}$ 	&	31.70	&	138.91 	&	2.13	$\pm$	0.03	&	5.30	$\pm$	0.10	&		-90.86 $\pm$ 0.04	& ---	\\
286746.514	&	13$_{1,12}$ -- 12$_{1,11}$ 	&	32.40	&	139.61 	&	2.53	$\pm$	0.02	&	5.53	$\pm$	0.06	&	-90.91	$\pm$	0.02 & ---	\\
306553.733 & 14$_{1,14}$ -- 13$_{1,13}$ & 39.70 & 153.62 & 2.13  $\pm$ 0.02 & 5.91 $\pm$ 0.07 & -90.86 $\pm$ 0.02  & --  \\
308798.184 & 14$_{1,13}$ -- 13$_{1,12}$ & 40.60 & 154.43 & 2.74  $\pm$ 0.02  & 5.83  $\pm$ 0.07  & -90.86 $\pm$ 0.02 & --  \\
330848.569 & 15$_{1,14}$ -- 14$_{1,13}$ & 50.10 & 170.31 & 4.64 $\pm$ 0.05 & 7.40 $\pm$ 0.10& -90.56 $\pm$ 0.04 & [5] \\
350333.059  & 16$_{1,16}$ -- 15$_{1,15}$ & 59.70 & 186.20 & 3.24 $\pm$ 0.04  & 5.84 $\pm$ 0.09 & -90.81 $\pm$ 0.03 & [6] \\
352897.581	&	16$_{1,15}$ -- 15$_{1,14}$ 	&	61.00 &	187.25	&	2.55	 $\pm$ 	0.03	&	5.25	$\pm$	0.08	&	-90.81	$\pm$	0.03 & ---	\\
\hline
175791.267	&	8$_{2,7}$ -- 7$_{2,6}$ 	&	6.65	&	208.25 	&	0.37	$\pm$	0.01	&	7.60	$\pm$	0.30	&	-92.20	$\pm$	0.10 & 	--- \\
175792.957	&	8$_{2,6}$ -- 7$_{2,5}$ 	&	6.65	&	208.25	&	0.41	$\pm$	0.03	&	7.90	$\pm$	0.60	&	-89.40	$\pm$	0.30 & 	--- \\
197762.939	&	9$_{2,8}$ -- 8$_{2,7}$ 	&	9.66	&	217.74	&	0.51	$\pm$	0.02	&	9.10	$\pm$	0.40	&	-92.50	$\pm$	0.20 & 	--- \\
197765.372	&	9$_{2,7}$ -- 8$_{2,6}$ 	&	9.66	&	217.74 	&	0.52	$\pm$	0.05	&	9.20	$\pm$	0.40	&	-88.8	$\pm$	0.10 & 	--- \\
219733.850$^\dagger$	&	10$_{2,9}$ -- 9$_{2,8}$ 	&	13.50	&	228.29	&	--  	&	-- 	&	--   & 	[3] \\
219737.193$^\dagger$	&	10$_{2,8}$ -- 9$_{2,7}$ 	&	13.50	&	228.29 	&	-- 	&	-- 	&	--  & 	[3] \\
241703.852$^a$	&	11$_{2,10}$ -- 10$_{2,9}$  	&	18.10	&	239.89 	&	--	          	&	     --           	&	--              & ---	\\
241708.312	&	11$_{2,9}$  -- 10$_{2,8}$  	&	18.10	&	239.89	&	0.67 $\pm$ 0.04	&	4.90 $\pm$ 0.40	&	-91.20 $\pm$ 0.10 & ---	\\
285640.923	&	13$_{2,12}$ -- 12$_{2,11}$ 	&	30.40	&	266.25 	&	0.63	$\pm$	0.05	&	6.40	$\pm$	0.60	&	-91.60	$\pm$	0.20 & ---	\\
285648.301	&	13$_{2,11}$ -- 12$_{2,10}$ 	&	30.40	&	266.25	&	0.53	$\pm$	0.04	&	6.20	$\pm$	0.60	&		-91.30 $\pm$	0.1 & ---	\\
307607.799	&	14$_{2,13}$ -- 13$_{2,12}$ 	&	38.20	&	281.01 &	0.64	$\pm$	0.02	&	7.70   	$\pm$ 0.40	&	-91.70	$\pm$	0.1 	& ---	\\
307617.020	&	14$_{2,12}$ -- 13$_{2,11}$ 	&	38.20	&	281.01 	&	0.47	$\pm$ 0.02		&	6.10   	$\pm$	0.30 	&	-90.70	$\pm$	0.10  & ---	\\
329573.452	&	15$_{2,14}$ -- 14$_{2,13}$ 	&	47.20	&	296.83 &	0.57	$\pm$	0.05	&	5.50	$\pm$	0.60	&	-91.30	$\pm$ 0.20	& ---	\\
329584.800	&	15$_{2,13}$ -- 14$_{2,12}$ 	&	47.20	&	296.83 	&	0.64	$\pm$ 0.08		&	7.00  	$\pm$	1.00  	&	-90.8	$\pm$	0.30 & ---	\\
351537.795	&	16$_{2,15}$ -- 15$_{2,14}$ 	&	57.50	&	313.70	&	0.72	$\pm$	0.07	&	5.80	$\pm$	0.70	&	-91.60	$\pm$ 0.20	& ---	\\
351551.573	&	16$_{2,14}$ -- 15$_{2,13}$ 	&	57.50	&	313.70	&	0.46	$\pm$ 0.04		&	5.70	$\pm$	0.70	&	-90.80	$\pm$	0.20 & ---	\\
\hline
\end{tabular}\\
Notes and labels: Columns 5--7 list Gaussian fit parameters with their respective uncertainties.$^\dagger$Weak and unresolved lines. $^a$Line likely affected by dominant emission of CH$_3$OH. NIST references, when available: [1] \citet{Arm84}; [2] \citet{Lor84}; [3] \citet{Sut85}; [4] \citet{Gre91}; [5] \citet{Sut91}; [6] \citet{Mac96}; [7] \citet{Jew89}.

\end{table*}

\begin{figure*}
    \centering
     \includegraphics[scale=0.53]{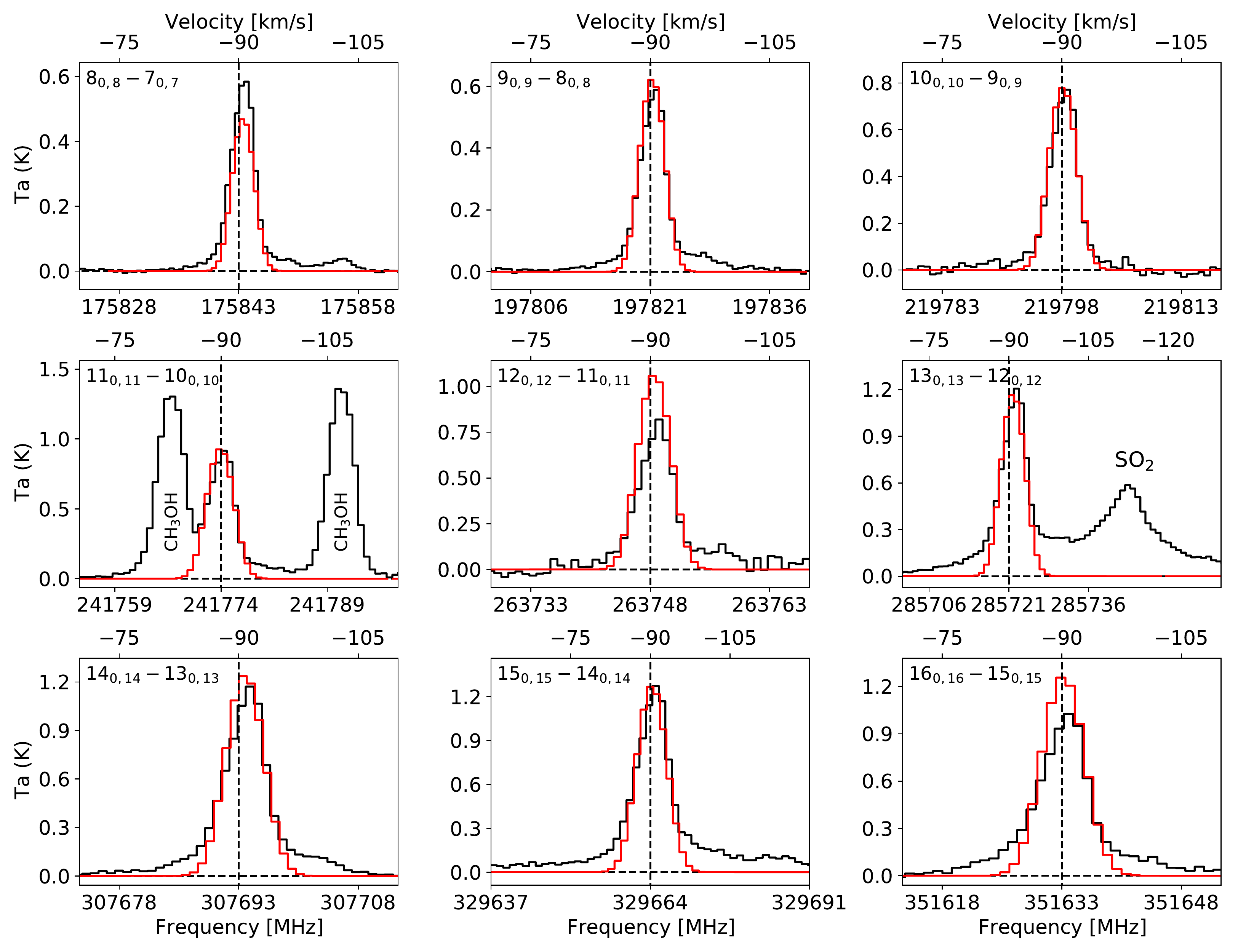}
    
     \caption{HNCO emission lines ($K_a$~=~0) and their LTE models, exhibited as black and red histograms, respectively. The line intensities, $y$-axes, are given in units of antenna temperature, the $x$-axes represent the frequency (bottom) and velocity (top). The horizontal and vertical dashed lines represent the baseline and the rest velocity of G331, respectively. The transitions are indicated in the top left corner of each plot. The spectral resolution was smoothed to 1 km s$^{-1}$.}
     \label{fig:k0}
\end{figure*}

\begin{figure*}
    \centering
     \includegraphics[scale=0.5]{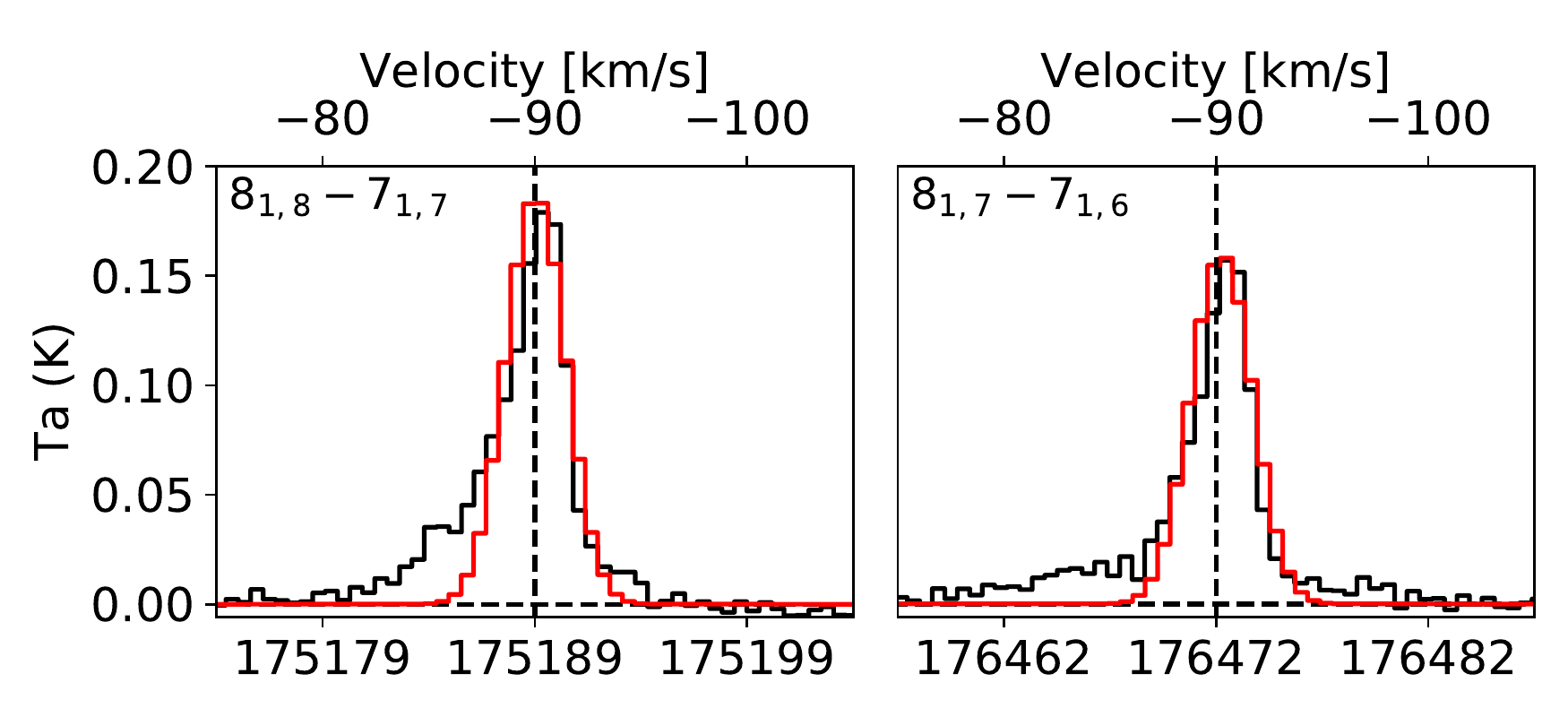}
     \includegraphics[scale=0.5]{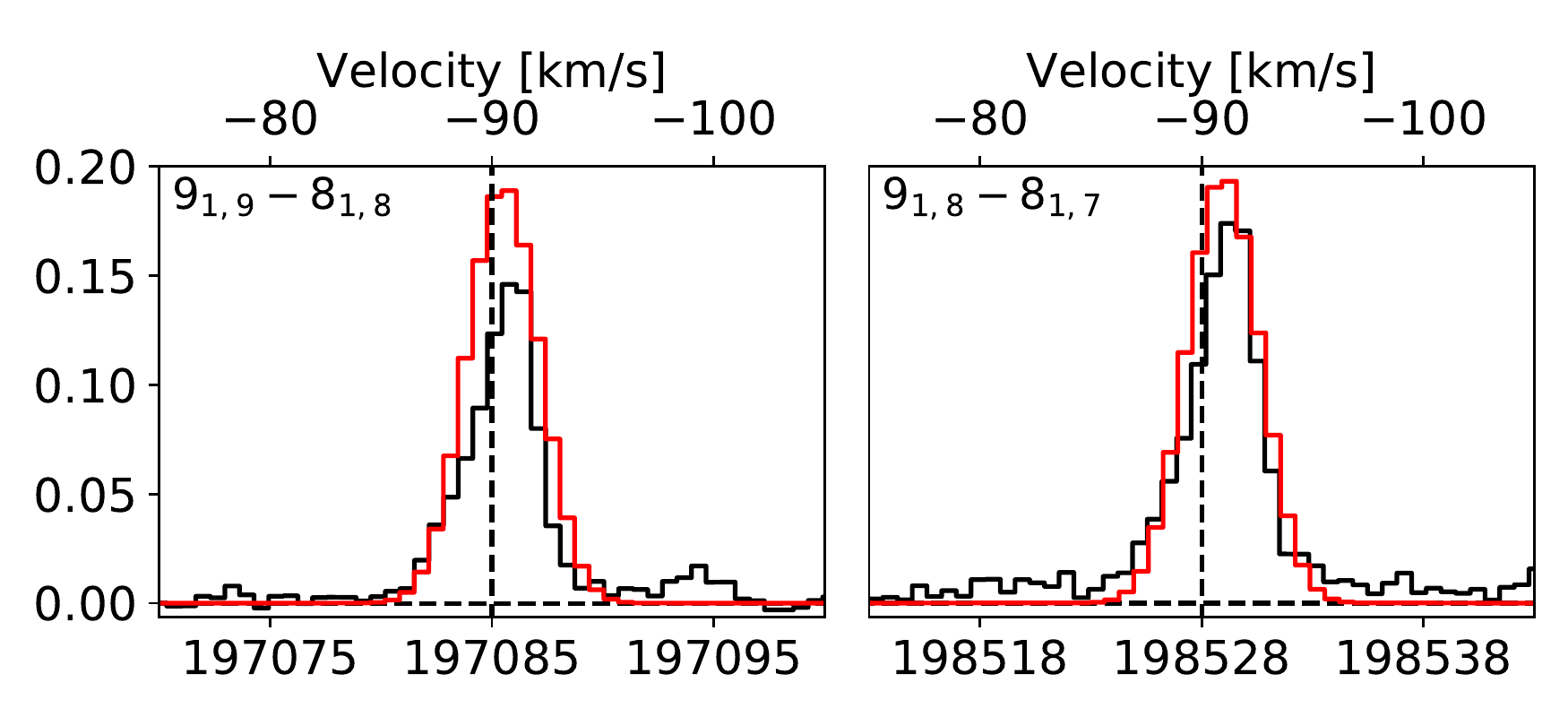}
     \includegraphics[scale=0.5]{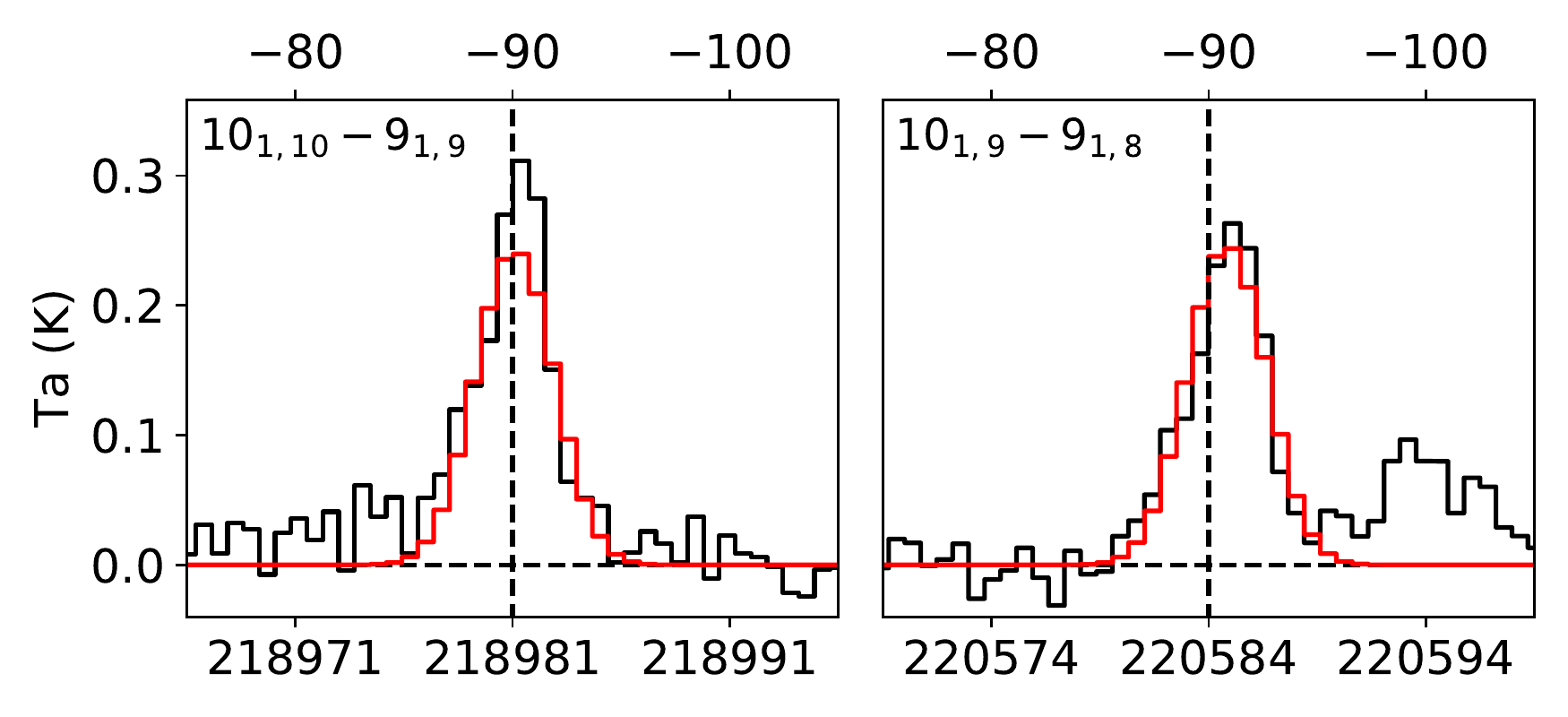}
     \includegraphics[scale=0.5]{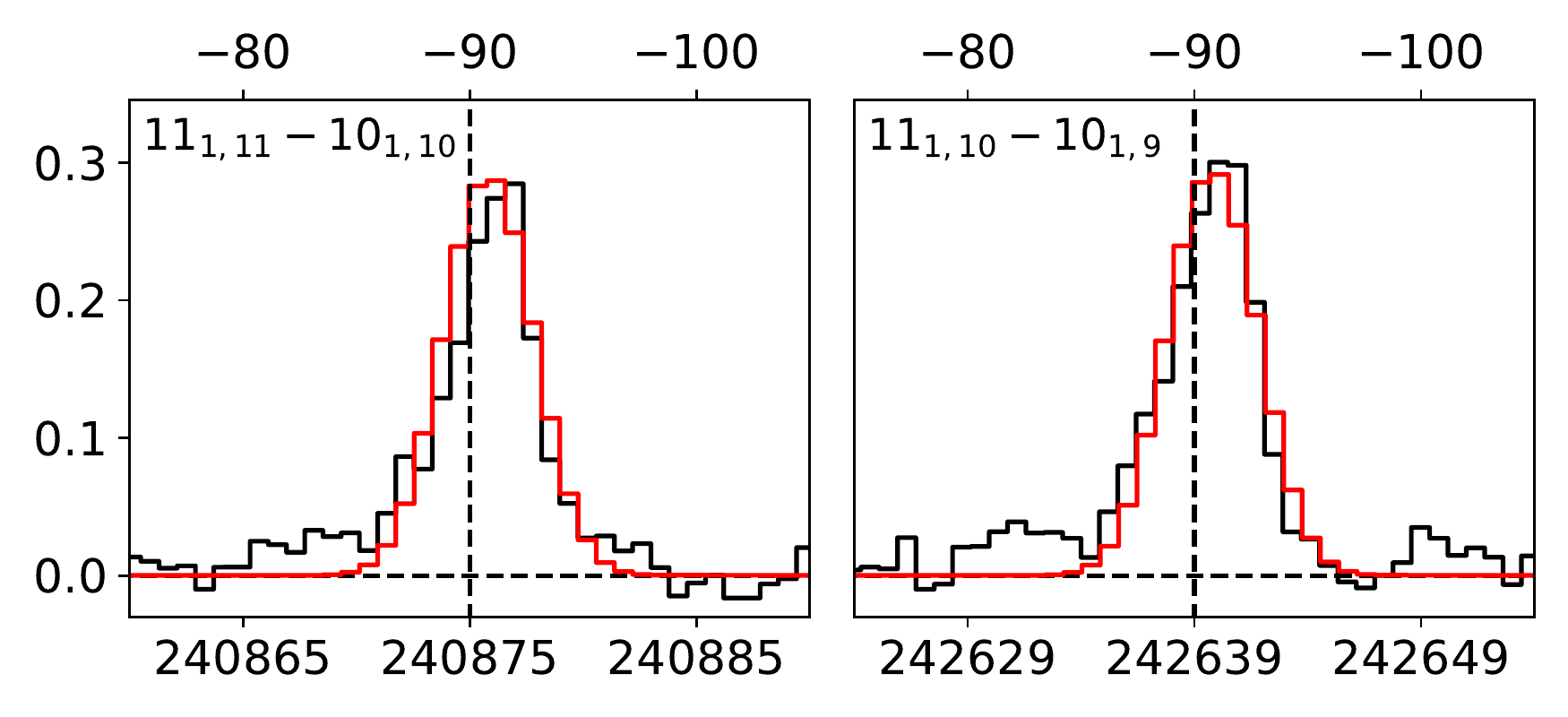}
     \includegraphics[scale=0.5]{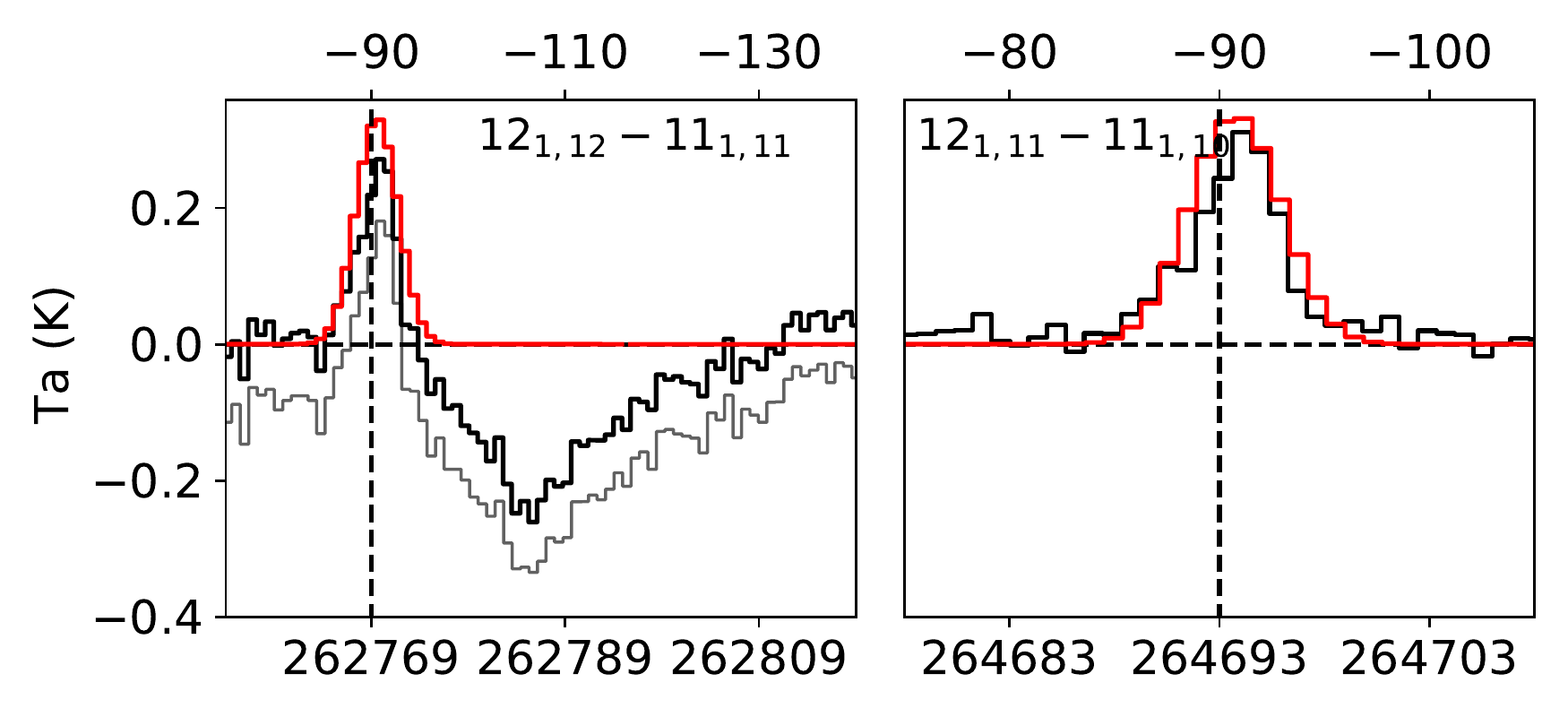}
     \includegraphics[scale=0.5]{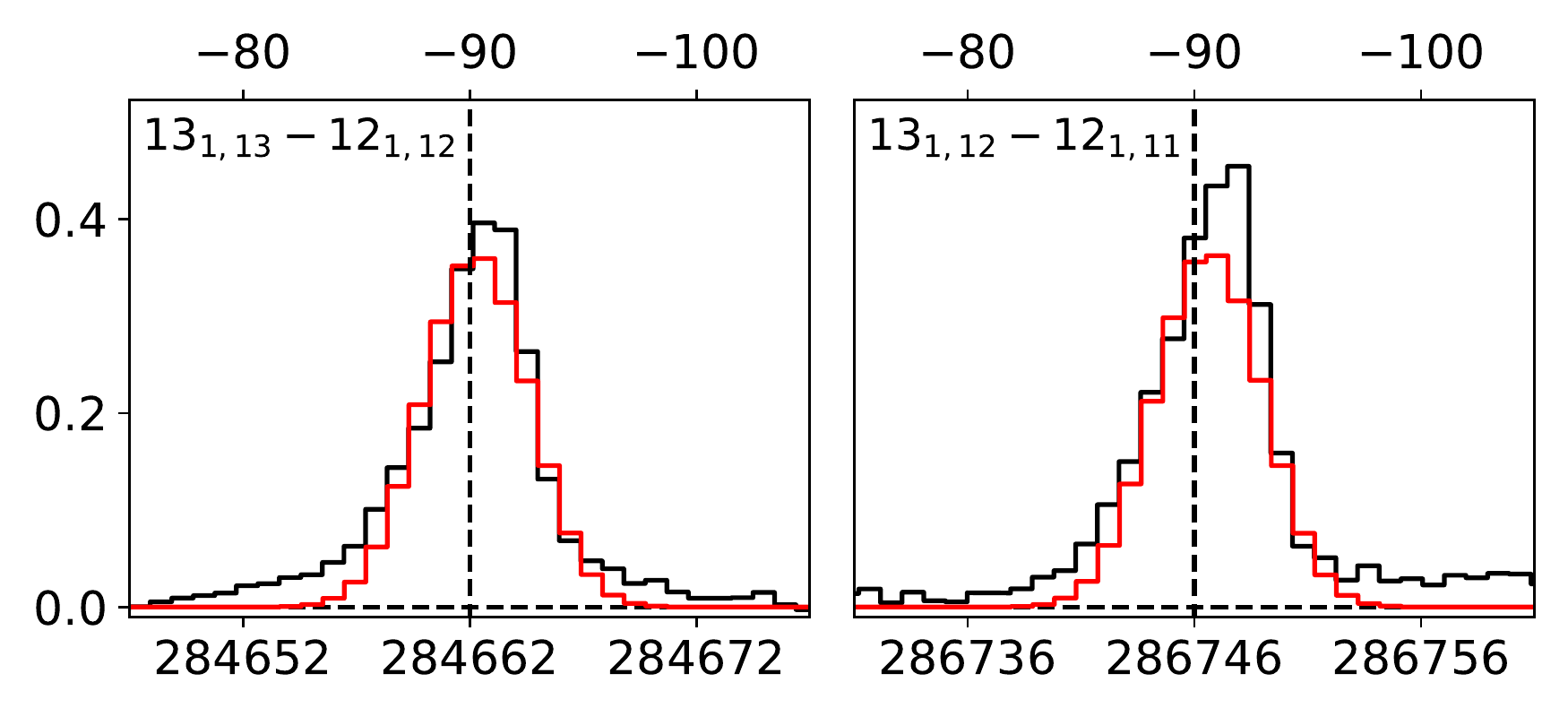}
     \includegraphics[scale=0.5]{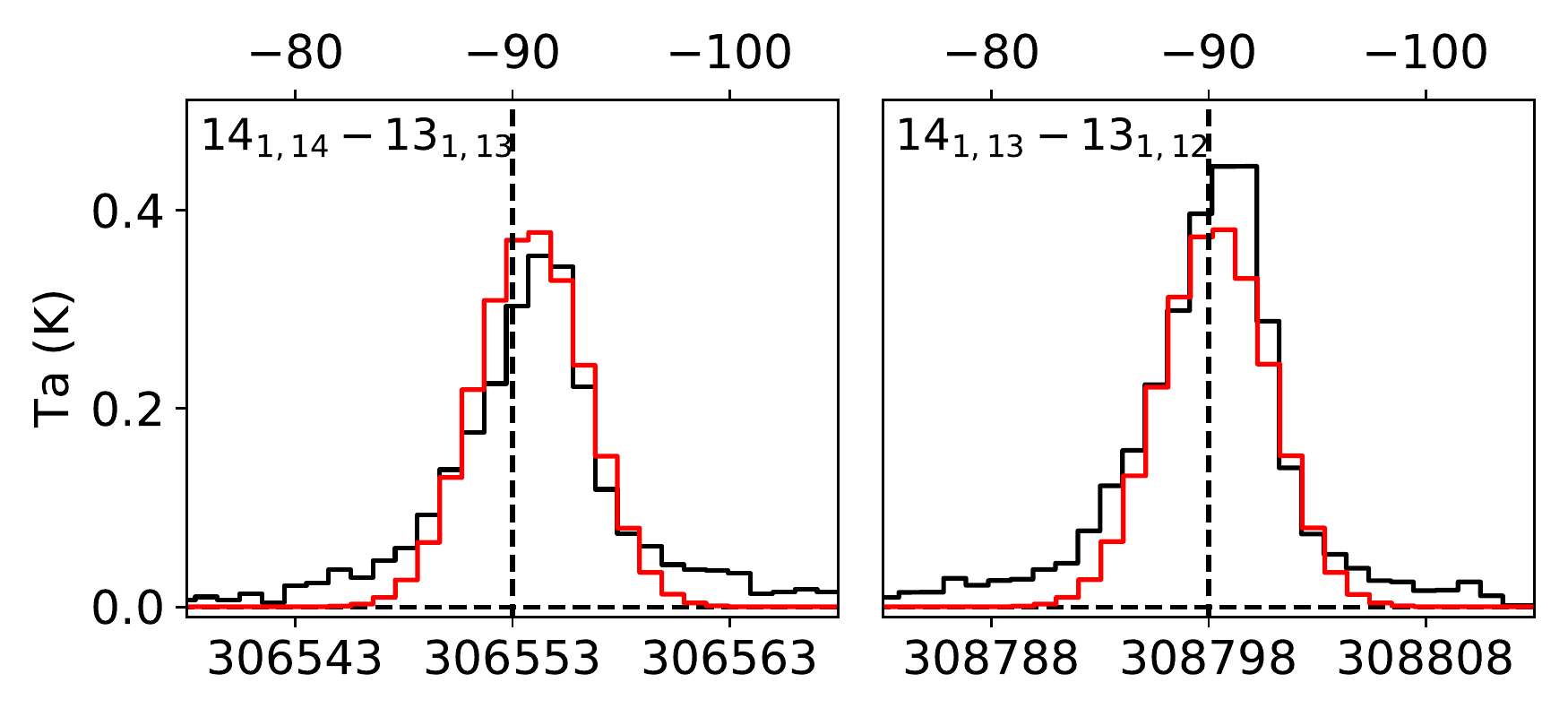}
     \includegraphics[scale=0.5]{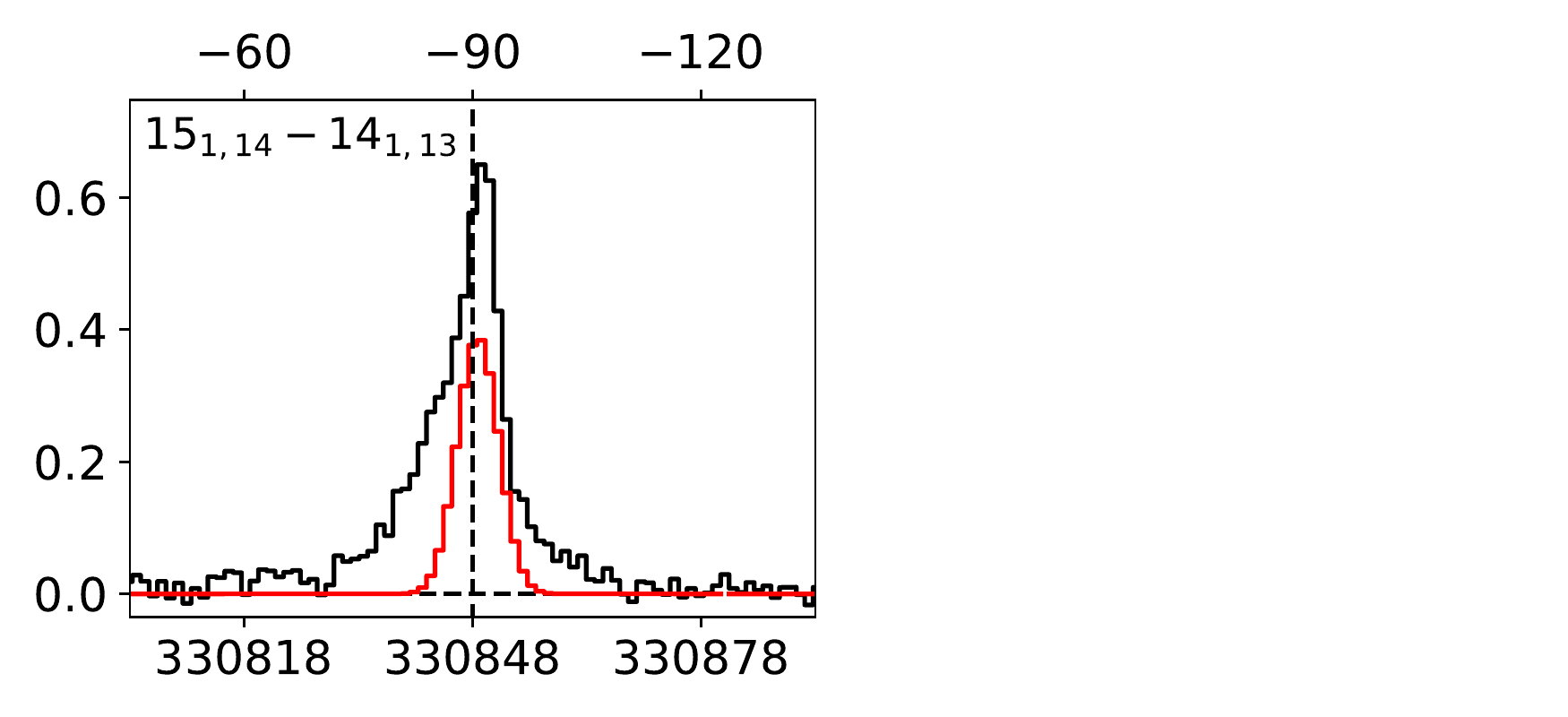}
     \includegraphics[scale=0.5]{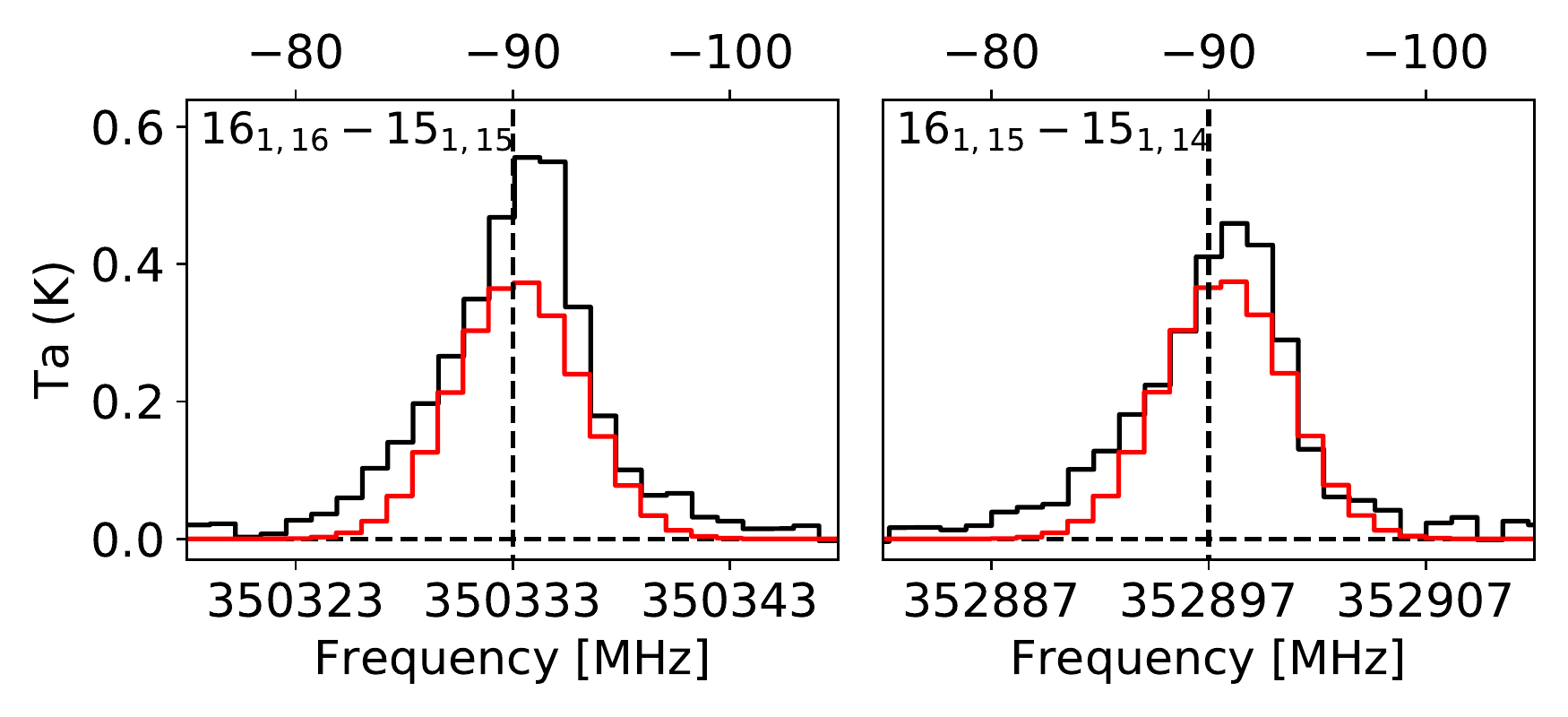}

     \caption{As in the caption of Fig.~\ref{fig:k0}. HNCO emission lines ($K_a$~=~1) and their LTE models. The transitions are indicated in the top  left corner of each plot. In the panel of the  12$_{1,12}$ -- 11$_{1,11}$ line at $\sim$ 262769~MHz, the black spectrum was adjusted from the grey one which is likely  compromised by an atmospheric issue at  $\sim$ 262785~MHz (see the text).}
     \label{fig:k1}
\end{figure*}

Nine lines of HNCO $K_a$~=~0 have been  detected  and  are exhibited in Fig.~\ref{fig:k0}. They have been  identified by selecting and superposing only HNCO $K_a$~=~0 transitions over the spectral survey. The line intensities of those transitions are the highest in comparison to the  $K_a$~=~1, 2 and even 3 ladders (Fig.~\ref{fig:k1},~\ref{fig:k2.1} and ~\ref{fig:HNCOk3}). Examining the spectra, an antenna temperature value around 0.6~K could differentiate the HNCO $K_a$ ladders,  while $K_a$~=~1, 2 and 3 exhibited $T_a \lesssim$~0.6~K. As a general perception in the observed $K_a$ ladders, the intensities seem to be in reasonably agreement with the line strengths (i.e. Einstein coefficients) found in spectroscopic databases (\S~\ref{sec:methods}), since they present  a slight increment as a function of the frequency. That tendency can also be appreciated in Table~\ref{tab:lines-v2}, where we present  the spectroscopic properties and estimated fluxes (i.e. integrated intensities) of the HNCO lines.

Among all the spectral lines, only two are likely  affected by blended or contaminant emission.Those are  the 11$_{0,11}$--10$_{0,10}$ and 13$_{0,13}$--12$_{0,12}$ transitions identified at 241774.032 and 285721.951~MHz, respectively (see Fig.~\ref{fig:k0}). For the first one, two neighbour lines of CH$_3$OH, appearing at 241767.247~MHz (5$_{-1,5}$ -- 4$_{-1,4}$ E) and 241791.367~MHz (5$_{0,5}$ -- 4$_{0,4}$, A), are likely responsible for the blending. Those transitions were reported in previous works by \citet{Lor84} in OMC-1. For the second one, it was found partially blended by a broad line likely associated to SO$_2$ v~=~0 (17$_{3,15}$ -- 17$_{2,16}$) at the frequency $\sim$ 285743~MHz.

Table~\ref{tab:lines-v2} lists the spectroscopic properties of all the HNCO transitions. The first part of that table presents the HNCO $K_a$~=~0 spectral characterisation. The lowest and highest $K_a$~=~0 transitions are  HNCO $8_{0,8}$--7$_{0,7}$ and 16$_{0,16}$--15$_{0,15}$ identified at the frequencies $\sim$~175843.695 and  351633.257~MHz, respectively. Likewise, the upper energy levels ($E_u$) and Einstein coefficients ($A_{ul}$) that govern those transitions are in the ranges $E_u \sim$~37.98--143.46~K and $A_{ul} \sim$~(7.43--61.3)$\times$10$^{-5}$~s$^{-1}$, respectively. Gaussian functions are used to fit parameters as fluxes, FWHM and $V_{lsr}$ values, all of them in units of K km~s$^{-1}$, km s$^{-1}$ and km~s$^{-1}$, respectively. As presented in Table~\ref{tab:lines-v2}, the rest frequencies of the lines are  well identified at $V_{lsr} \thickapprox$-90~km~s$^{-1}$ and the linewidths present  a low dispersion of FWHM values, between $\sim$~5.5 and 6.7~km~s$^{-1}$. Fluxes are  found to be in the range $\sim$~3.63--8.02~K km~s$^{-1}$, with the highest values for the lines at highest frequencies, according to APEX SHeFI~2. For the mentioned range, the lowest and highest flux values correspond  to HNCO 9$_{0,9}$--8$_{0,8}$ and 15$_{0,15}$--14$_{0,14}$ at 197821.461~MHz and 329664.367~MHz, respectively.

\subsubsection*{HNCO $K_a$~=~1}

Seventeen lines of HNCO $K_a$~=~1 have been observed as spectral pairs within the same rotational transitions. Their spectra are exhibited in Fig.~\ref{fig:k1}. As we have already done  for $K_a$~=~0 ladder, $K_a$~=~1 transitions have been chosen and superposed over the spectral survey. The intensities and fluxes of the $K_a$~=~1 ladders are lower than those with $K_a$~=~0. The $K_a$~=~1 lines have been  observed across a wide frequency interval, from $\sim$175189.027~MHz to 352897.581~MHz, and they are  stronger as the frequency increases. As we have highlighted for the $K_a$~=~0 ladder, such tendency can be corroborated through the spectrum  and the fitted  parameters listed in the middle part of Table~\ref{tab:lines-v2}. Concerning to problems affecting the spectral emission, almost all the lines are clean or free of contamination.

Two aspects can be highlighted from the spectra exhibited in Fig.~\ref{fig:k1}. On the one hand, the spectral profiles are not totally symmetrical, exhibiting a tail extended towards low frequency values, in some cases. On the other hand, regarding  contamination or problems in the spectra, the HNCO 10$_{1,9}$--9$_{1,8}$ line at 220584.751~MHz presents a neighbour unidentified emission that could be CH$_3$CN 12$_6$-11$_6$ at 220594.431~MHz, which was previously reported in OMC-1 \citep{Lor84}. Moreover, in the panel of the HNCO 12$_{1,12}$--11$_{1,11}$ line at 262769.477~MHz, a feature in absorption appears  at $\sim$~262787~MHz affecting the whole spectrum. Such absorption is likely associated with water vapour as it has been discussed in studies about the atmospheric transparency at Chajnantor.\footnote{\url{https://almascience.eso.org/about-alma/atmosphere-model}} 

\begin{figure}
    \centering
    \includegraphics[width=\columnwidth,keepaspectratio]{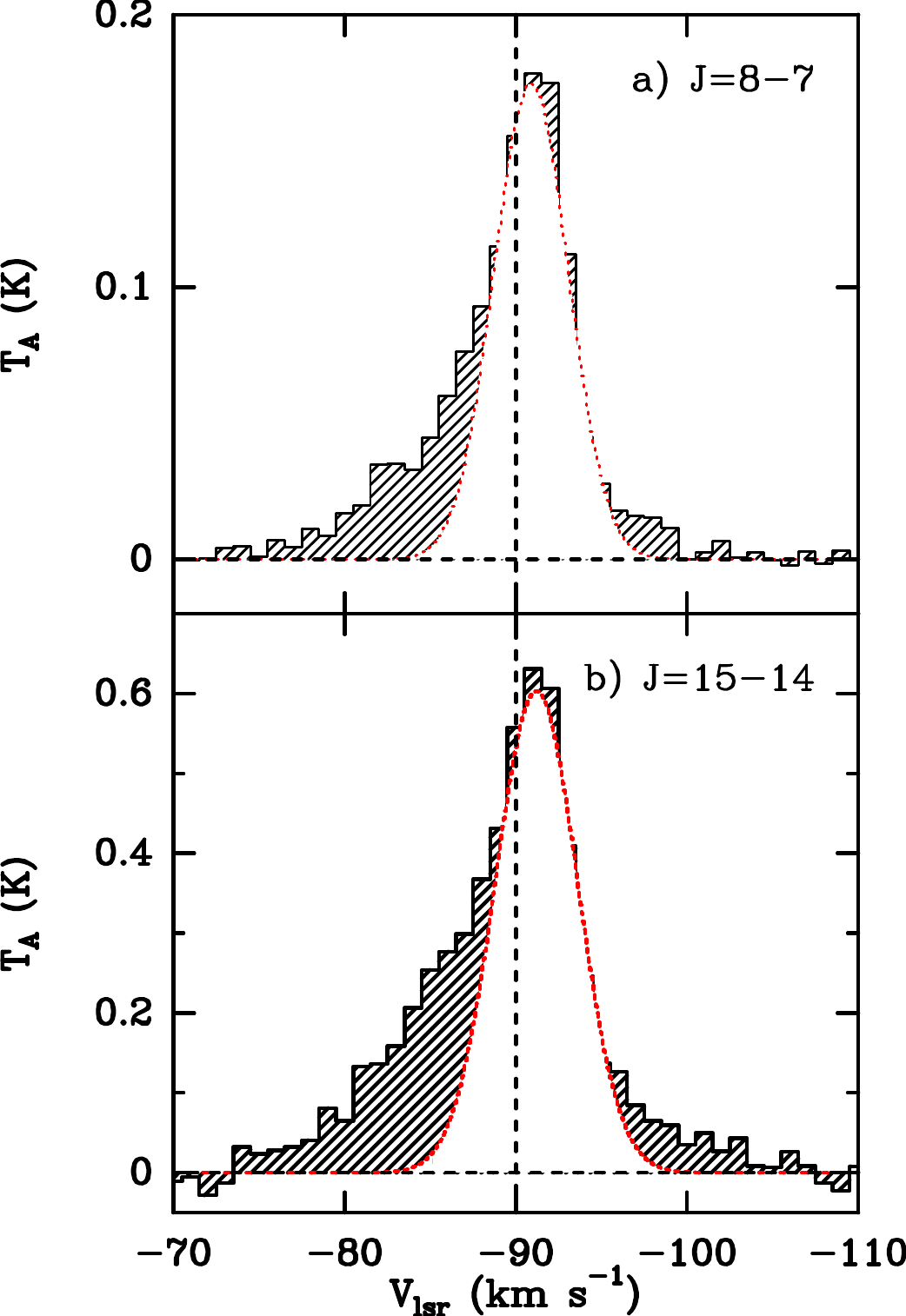}    
    \caption{Spectral wings observed in the (a) HNCO 8$_{1,8}$--7$_{1,7}$ and (b) 15$_{1,14}$--14$_{1,13}$  transitions identified at the rest frequencies $\sim$~175189.027 and 330848.569~MHz, respectively. The diagonal patterns depict the area under the spectra from $-$70 to $-$110~km~s$^{-1}$. The red dotted lines  depict the areas under the Gaussian fits.}
    \label{fig:i4}
\end{figure}

Concerning the line profiles, the spectral tail mentioned above has also been  evidenced in HNCO lines detected in OH 231.8+4.2, which is an oxygen-rich circumstellar envelope around an intermediate-mass evolved star also harbouring a bipolar molecular outflow \citep{Prieto15}. So we speculate that such asymmetry in the lines could be due to outflow activity. In addition, it is also worth mentioning that such tail is more notorious in some transitions than others. It also seems to be better traced by the ladder $K_a$~=~1 than $K_a$~=~0. In Fig.~\ref{fig:i4},  we emphasise  the observed tails in the HNCO 8$_{1,8}$--7$_{1,7}$  and 15$_{1,14}$--14$_{1,13}$  lines, detected at 175189.027~MHz and 330848.569~MHz, respectively, since both lines exhibit remarkably well the one-side tail extended up to -70~km~s$^{-1}$. Comparing both lines, we note that the second one is more intense and stronger by a factor $\sim$~5 if we take, for instance, the ratio from their fluxes. Although other studies found spectral profiles of HNCO \citep{Prieto15} similar to those observed here, we took into account a scenario of possible contamination affecting the HNCO emission (Fig.~\ref{fig:i4}). However, such scenario is unlikely due to the rare species and/or transitions theoretically predicted. In the case of the HNCO 8$_{1,8}$--7$_{1,7}$ line detected at 175189.027~MHz, transitions of e.g. CH$_2$ND, C$_2$H$_5$OH and i-C$_3$H$_7$CN are predicted at $\sim$~175185.19, 175188.99 and 175189.44~MHz. Regarding the HNCO 15$_{1,14}$--14$_{1,13}$ line detected at 330848.569~MHz, which was catalogued in Sgr B2(N) by \citet{Sut91}, transitions of e.g. HCOCN  and $^{15}$NH$_2$HCO at $\sim$~330845.85 and 330851.20~MHz, respectively, could be possible contaminants. Thus, the observed  spectral profiles of HNCO seem to be much better explained by the outflow activity of G331.

In spite of the spectral asymmetries shown in Fig.~\ref{fig:i4}, a relative percentage was estimated by considering the area under the Gaussian fits with respect to the whole area integrated within $-70~\leq V_{lsr}$(km~s$^{-1}$)$~\leq -110$. Thus, the areas under the Gaussian fits represent up to $\thickapprox$~80~per~cent of the total area under the spectra. Although these lines have different intensities and fluxes (Fig.~\ref{fig:i4}b), the percentages are similar as both lines have similar spectral tails.   \citet{Prieto15} also observed this wing, through spectral lines of HNCO, but in outflows of AGB stars.

\subsubsection*{HNCO $K_a$~=~2}

\begin{figure*}
     \centering
     \includegraphics[scale=0.67]{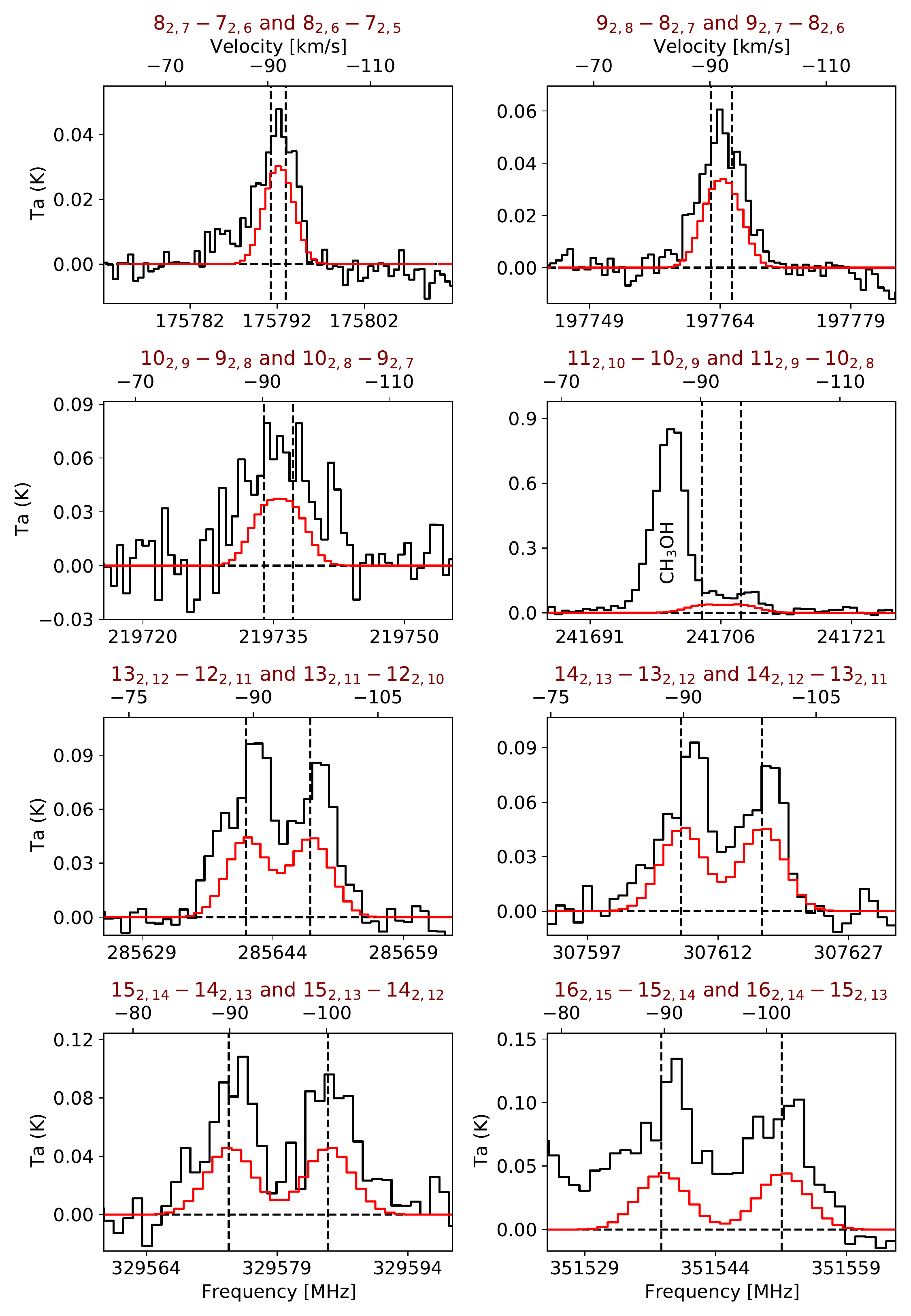}
     \caption{As in the caption of Fig.~\ref{fig:k0}. HNCO emission lines ($K_a$~=~2) and their LTE models. The transitions are indicated in the top  left corner of each plot.}
     \label{fig:k2.1}
 \end{figure*}

Sixteen transitions of HNCO $K_a$~=~2 have been  identified across the survey, they appear  as spectral pairs partially resolved.  The spectral lines are exhibited in Fig.~\ref{fig:k2.1}, where it can be noted that their emission are noisier and weaker than the previously analysed $K_a$ ladders; those were aspects that put constraints on the line identification of the HNCO $K_a$~=~2 transitions. Likewise, it can be observed that the intensity of the lines increases with the frequency. In addition,  the spectral pairs are more spaced  as the frequency increases.  For instance, by comparing the two panels centred at 175792~MHz and 329579~MHz of Fig.~\ref{fig:k2.1}, it can be noted that the spectral pair of the second one, 15$_{2,14}$--14$_{2,13}$ and 15$_{2,13}$--14$_{2,12}$, is better resolved than the first one, 8$_{2,7}$--7$_{2,6}$ and 8$_{2,6}$--7$_{2,5}$.  
Apart from the  HNCO blending mentioned above, only the $K_a$~=~2 transitions at $\sim$~241706~MHz  are affected by a neighbour and intense line of CH$_3$OH (5$_{0,5}$--4$_{0,4}$, E) identified at $\sim$~241700~MHz (see Fig.~\ref{fig:k2.1}). Concerning the line profiles, the lines do not exhibit the spectral tails like those highly evidenced in the HNCO $K_a$~=~1 ladder. 

In the lowest part of Table~\ref{tab:lines-v2}, we present the parameters of the observed HNCO $K_a$~=~2 transitions. Summarising, the lowest and highest transitions are  identified at the rest frequencies 175791.267 and 351551.573~MHz, respectively. Their upper energy levels, which  are in the range $E_u \sim$~208.25--313.70~K, are the highest in comparison with the other $K_a$ ladders. Concerning  the Einstein coefficients, they are in the range $A_{ul} \sim$ (6.65--57.5)$\times$10$^{-5}$~s$^{-1}$,  similar to those of the other $K_a$ ladders. In order to quantify the spectral emission, Gaussian functions have been adjusted to estimate the line parameters, whose results are presented in  Table~\ref{tab:lines-v2}. Since the lines are partially blended, the estimation of the Gaussian fits has been  difficult, mainly for the  lowest frequency transition pairs observed at $\sim$~175792, 197764, 219735 and 241706~MHz. In addition to that problem, and due to the weakness of the emission, the resultant Gaussian fits are not as good as those for the previously analysed $K_a$ ladders; as a consequence, the LTE analysis of the HNCO $K_a$~=~2 lines, e.g. the match between models and observations (Fig.~\ref{fig:k2.1}), might not be as reliable as those obtained from the strongest $K$ ladders observed in this work (Figs.~\ref{fig:k0} and \ref{fig:k1}). Regarding the rest frequencies, they were found more dispersed with respect to the systemic velocity, approximately between -88 and -92~km~s$^{-1}$. Intensities and fluxes were found to be as the weakest among all the $K_a$ ladders, see Table~\ref{tab:lines-v2}. The transitions 10$_{2,9}$--9$_{2,8}$, 10$_{2,8}$--9$_{2,7}$ and 11$_{2,10}$--10$_{2,9}$ were unable to obtain reliable Gaussian fits. 

\subsubsection*{HNCO $K_a$~=~3}

\begin{figure}
\includegraphics[width=\columnwidth,keepaspectratio]{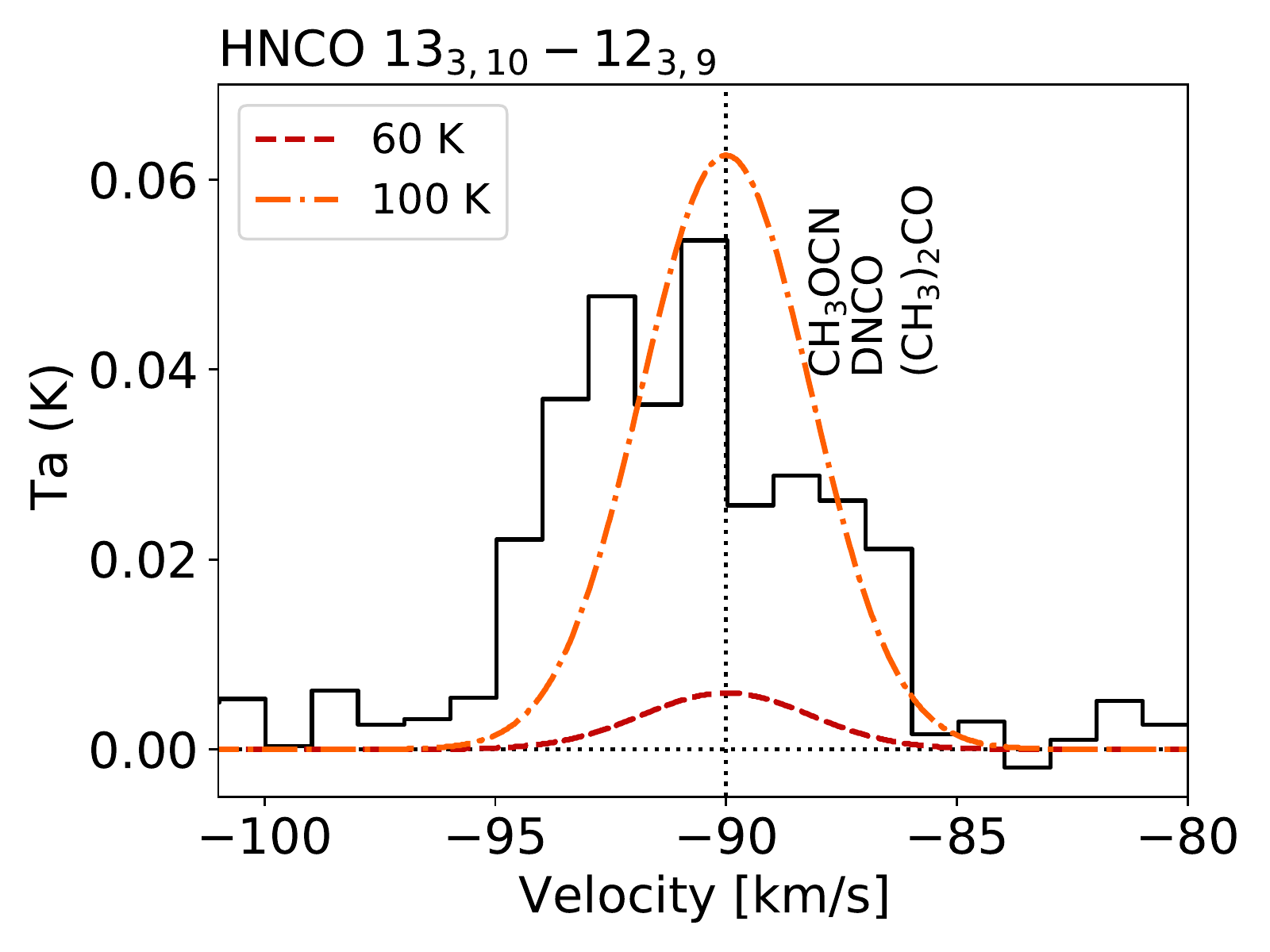}
\caption{Tentative detection of the HNCO line (13$_{3,10} - 12_{3,9}$) at the rest frequency 285541.575~MHz. The baseline and the source's velocity are represented by the dotted horizontal and vertical lines, respectively. The LTE models at 60~K and 100~K, displayed as red dashed and orange dot-dashed lines, respectively, are to illustrate the rotational diagram result and a hypothetical higher excitation condition, respectively. There is a fainter emission next to the HNCO line where we indicated other candidate molecules.}
\label{fig:HNCOk3}
\end{figure}

A subtle evidence for a HNCO $K_a$~=~3 line has been  found. The HNCO 13$_{3,10} - 12_{3,9}$ transition is  identified at the rest frequency 285541.575~MHz, and is exhibited in Fig.~\ref{fig:HNCOk3}. As expected, the spectral line is weak and has the highest upper energy level identified in this work, $E_u$~=~470.91~K, with $A_{ul}$~=~2.78$~\times~$10$^{-4}$~s$^{-1}$. As a consequence, such emission would suggest a higher excitation condition than that found from the other HNCO $K_a$~=~0, 1 and 2 ladders. As can be noted by the models at two different excitation temperatures displayed in Fig.~\ref{fig:HNCOk3}, the highest temperature does reproduce better the line profile.  Analysing other species for the peak, CH$_3$OCN at 285539.098~MHz, DNCO at 285539.531~MHz or acetone (CH$_3$)$_2$CO at 285542.349~MHz could be candidate transitions.  The predicted position in the spectrum for those species are also indicated in Fig.~\ref{fig:HNCOk3}.

\subsection{LTE analysis}

In order to estimate the physical conditions traced by HNCO in G331 we constructed rotational diagrams. They are an important method to infer source properties from molecular line emissions  \citep[see][for a complete description and discussion]{gol1999}. They basically consist of a plot of the column density per statistical weight, for a given number of molecular rotational energy levels, as a function of their  energies above the ground state. More specifically, if LTE is assumed for the source, the level populations can be represented by the Boltzmann distribution and the rotational diagram is described by Eq.~\ref{eq:rot-diag}.  Moreover, if the emission lines can also be considered as optically thin, the plotted function is a straight line with a slope defined by $1/T_{ex}$  from which is possible to derive the excitation  temperature of the source where it is the molecule, that should be equal to the kinetic temperature of the gas in LTE conditions. Thus,

\begin{equation}
    \ln\left ( \frac{N_{u}}{g_u} \right )= \ln\left( \frac{3k W}{8\pi^3\nu S_{ul}\mu^2 g_u} \right) = \ln\left ( \frac{N}{Q} \right )-\frac{E_u}{k \, T_{ex}}
    \label{eq:rot-diag}
\end{equation}

\noindent where $N_u$ is the column density of the upper level, $g_u$ the degeneracy of the upper level,  $k$  is the Boltzmann constant, $W$ the integrated intensity of the transition, $\nu$ the rest frequency of the line, $S_{ul}$ is the line strength of the transition, $\mu$ is the dipolar moment,  $N$ is the total column density, $Q$ is the internal partition function of the molecule, and $E_u$ the upper level energy of the transition.

For the construction of the rotational diagram, it is  also considered a beam dilution factor associated with a point-like emitter region. The correction can be performed by the addition of the term ln~($\Delta\Omega_a$/$\Delta\Omega_s$) on the right-hand side of Eq.~(\ref{eq:rot-diag}) \citep{gol1999}, where $\Delta\Omega_a$ is the antenna solid angle and $\Delta\Omega_s$ is the source solid angle. This ratio correlates the subtended angle of the source with the solid angle of the antenna beam. For the HNCO emission, observed through several spectral lines, we adopt a hypothetical source size of 5~arcsec  to evaluate the beam dilution effects,   although, under the hypothesis that the emission fills the antenna beam, a rotational diagram solution was also constructed considering 30~arcsec. The LTE treatment of HNCO was motivated by radiative analyses of HC$_3$N in G331, although the physico-chemical origin of both molecules is different. For instance, in previous works, various lines of HC$_3$N, including its $^{13}$C isotopologues, were analysed with APEX \citep{Duronea19}, as well as mapped with ALMA via the HC$_3$N $J$~=~38--37 transition \citep{Hervias19}. Furthermore, in the case of optically thick transitions, the optical depth correction can be performed  by the correction factor $C_{\tau}$ (Eq.~\ref{eq:ctau}) as shown in Eq.~\ref{eq:nthick} \citep[as derived in][]{gol1999}:

\begin{equation}
    N_{u,thick}=N_{u,thin} \times C_\tau
\label{eq:nthick}    
\end{equation}

\begin{equation}
    C_\tau = \frac{\tau}{1-e^{-\tau}} 
\label{eq:ctau}    
\end{equation}

\noindent where $\tau$ is the optical depth.

\begin{figure}
    \centering
    \includegraphics[width=\columnwidth,keepaspectratio]{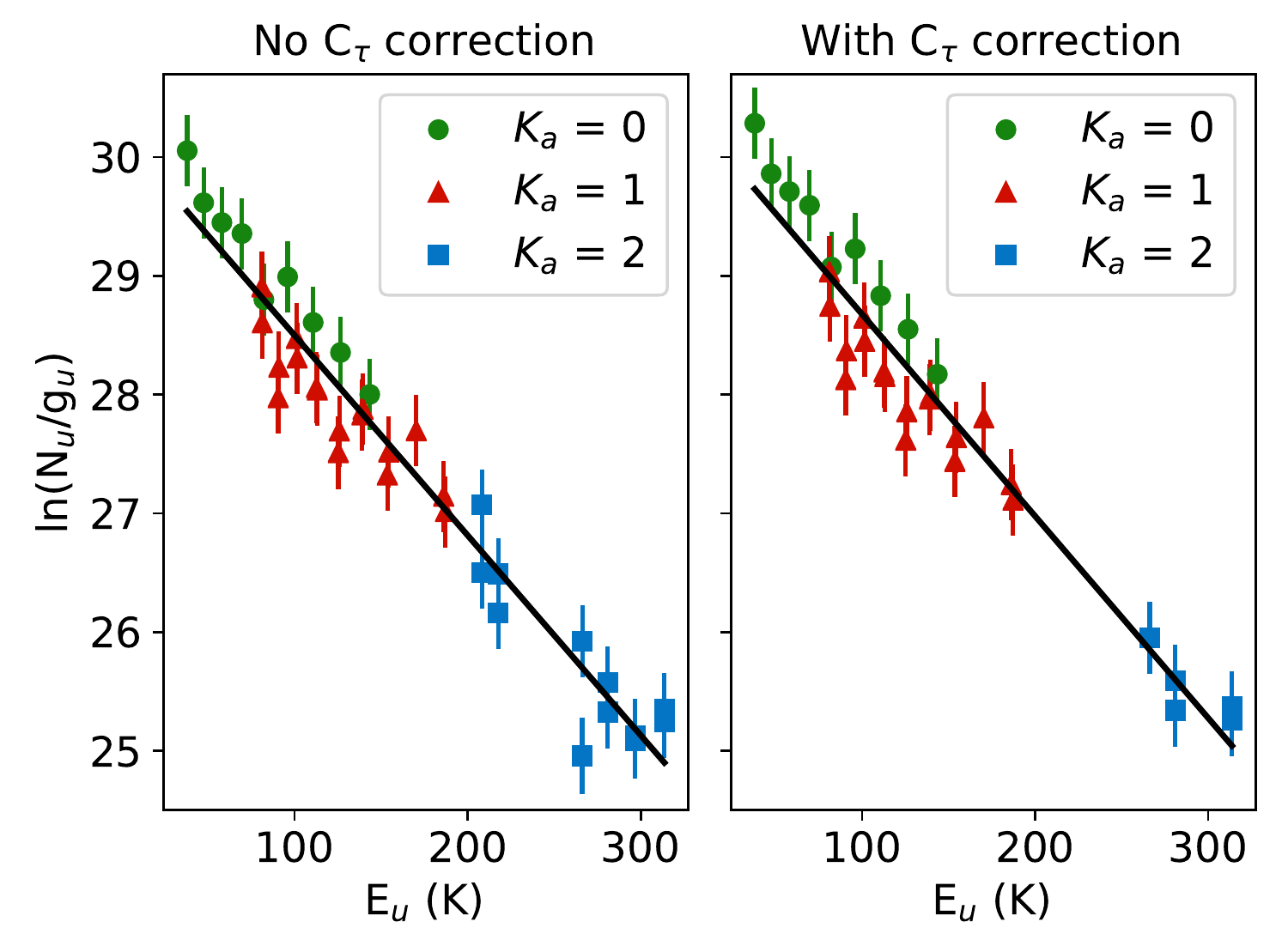}
    \caption{Rotational diagrams of HNCO constructed applying opacity correction (right panel) and without opacity correction (left panel). In both cases, a source size of 5~arcsec was adopted. In second plot, only the lines with no blended/contaminant emission are considered. The best linear fit for each display yields, respectively,  $T_{ex}$ = 59.4 $\pm$ 2.3~K and $N$ = (3.1 $\pm$ 0.4) $\times$ 10$^{15}$~cm$^{-2}$ ($\chi^2_{red}$=1.12); and $T_{ex}$= 58.8 $\pm$ 2.7~K and $N$ =  (3.7 $\pm$ 0.5) $\times$ 10$^{15}$~cm$^{-2}$ ($\chi^2_{red}$=1.21). The set of data with different $K_a$ values are also represented in the plots.}
    \label{fig:rd3}
\end{figure}

The rotational diagrams with and without  opacity correction (\ref{eq:nthick}) are presented in Fig.~\ref{fig:rd3}.  Regarding the rotational diagram corrected by the opacity, it considers only the non-blended transitions of HNCO because the opacity correction does not converge on blended transitions, like various of those belonging to the  $K_a$~=~2 ladder and identified at lowest frequency values (e.g. the first panels of Fig.~\ref{fig:k2.1}). The best linear fit for the diagrams gave $N$(HNCO) = (3.1 $\pm$ 0.4) $\times$ 10$^{15}$~cm$^{-2}$ and $T_{ex}$= 59.4 $\pm$ 2.3~K, without the opacity correction; and  $N$(HNCO) =  (3.7 $\pm$ 0.5) $\times$ 10$^{15}$~cm$^{-2}$ and $T_{ex}$= 58.8 $\pm$ 2.7~K, by applying the opacity correction. The reduced $\chi^2$ values for each fit are 1.12 and  1.21, respectively. Assuming that the HNCO emission is extended, optically thin and that fills approximately the antenna beam, we evaluated a rotational diagram adopting 30~arcsec, the best linear fit ($\chi^2_{red}$=1.98) gave $N$(HNCO) = (1.4 $\pm$ 0.2) $\times$ 10$^{14}$~cm$^{-2}$ and $T_{ex}$= 65.5 $\pm$ 3.7~K. With similar results, such temperatures around 60~K might indicate that HNCO molecules are concentrated in external and colder regions of the core, according to the proposed shell-like structure of G331 \citep{Merello13a,Hervias19,Duronea19}.

Moreover, \citet{Churchwell1986}  indicated that HNCO transitions with  $E_u >$~40~K  require an excitation temperature of $\sim$~70~K and also that radiative processes are responsible for the excitation of HNCO in Sgr B2, as already mentioned. This suggests that HNCO can be considered as a good probe of the far-infrared radiation field but not of gas properties such as density and kinetic temperature. The only common transition with our work is 10$_{0,10}$--9$_{0,9}$, the others are at lower frequencies \citep{Churchwell1986}. In this sense, according to the similar excitation temperature of $\sim$~60~K obtained with the rotational diagrams, the same trend appears to be maintained at higher HNCO transition levels. This may imply that radiative processes, rather than collisional mechanisms, could also dominate the excitation of HNCO molecules in G331. In this article, it is reported for first time a multiline analysis of HNCO, covering transitions within the energy interval $E_u \thickapprox$ 37--314~K, towards a massive hot molecular core/outflow like G331. Recently, \citet{He2021} studied the spatial distribution of HNCO 4$_{0,4}$--4$_{0,3}$, SiO 2--1 and HC$_3$N 10--9 in a sample of southern massive star-forming regions, discussing correlations between the morphology of the gas and dust emission. In an earlier work, \citet{Martin2008} proposed how HNCO and CS could be useful molecules to study the influence of shocks and/or the radiation field in nuclear regions of galaxies. Thus, there are follow-up questions that warrant further studies to have a deeper understanding on the role that the outflow activity, the physical conditions of the dust and collisional mechanisms can play into the formation, distribution and excitation of HNCO gas  in massive protostellar objects like G331.

\section{Discussion}
\label{sec:disc}

\subsection{Abundances: comparison with other sources}

From the column densities obtained with the rotational diagram, assuming a  source size of 5~arcsec, it is possible to estimate the abundance of HNCO if the column density of H$_2$ is also known. The  most recent values of $N$(H$_2$) derived for G331 are 9.7  $\times$ 10$^{23}$~cm$^{-2}$, obtained from the H$^{13}$CO$^+$/H$_2$ ratio of Orion KL, and 2.7 $\times$ 10$^{23}$~cm$^{-2}$, estimated by the superficial density from the continuum emission at 1.2~mm in G331 \citep[][and references therein]{Duronea19}. Therefore, by taking the  column density of HNCO corrected by the optical depth,  $N$(HNCO)=(3.7$\pm$0.5) $\times$ 10$^{15}$~cm$^{-2}$, and adopting H$_2$ column densities values of 2.7 $\times$ 10$^{23}$ and 9.7 $\times$ 10$^{23}$~cm$^{-2}$, the relative abundances of HNCO are estimated to range from (3.8 $\pm$ 0.5) $\times$ 10$^{-9}$ to (1.4 $\pm$ 0.2) $\times$ 10$^{-8}$. 

\begin{figure}
    \centering
    \includegraphics[width=\hsize]{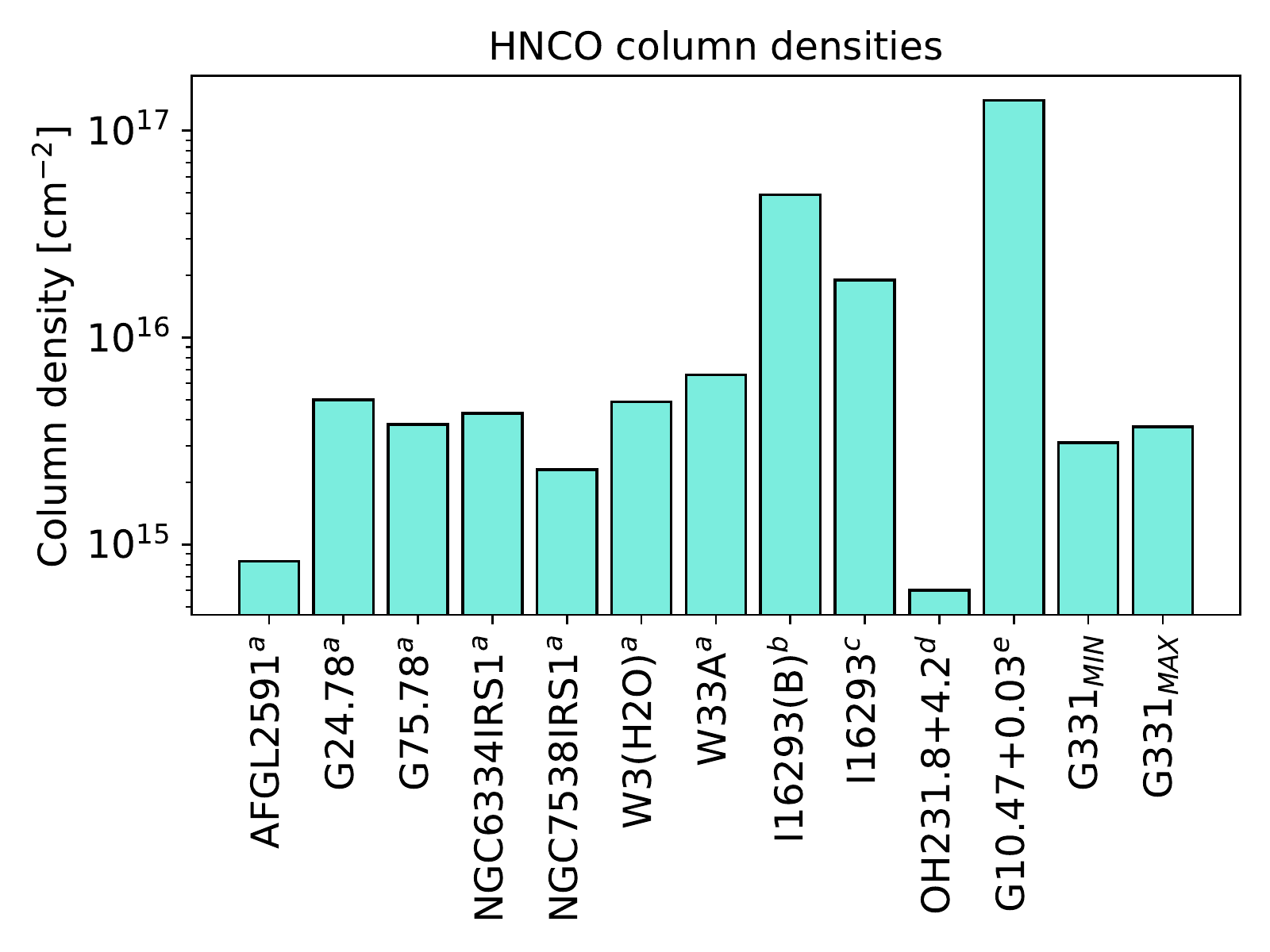}
    \includegraphics[width=\hsize]{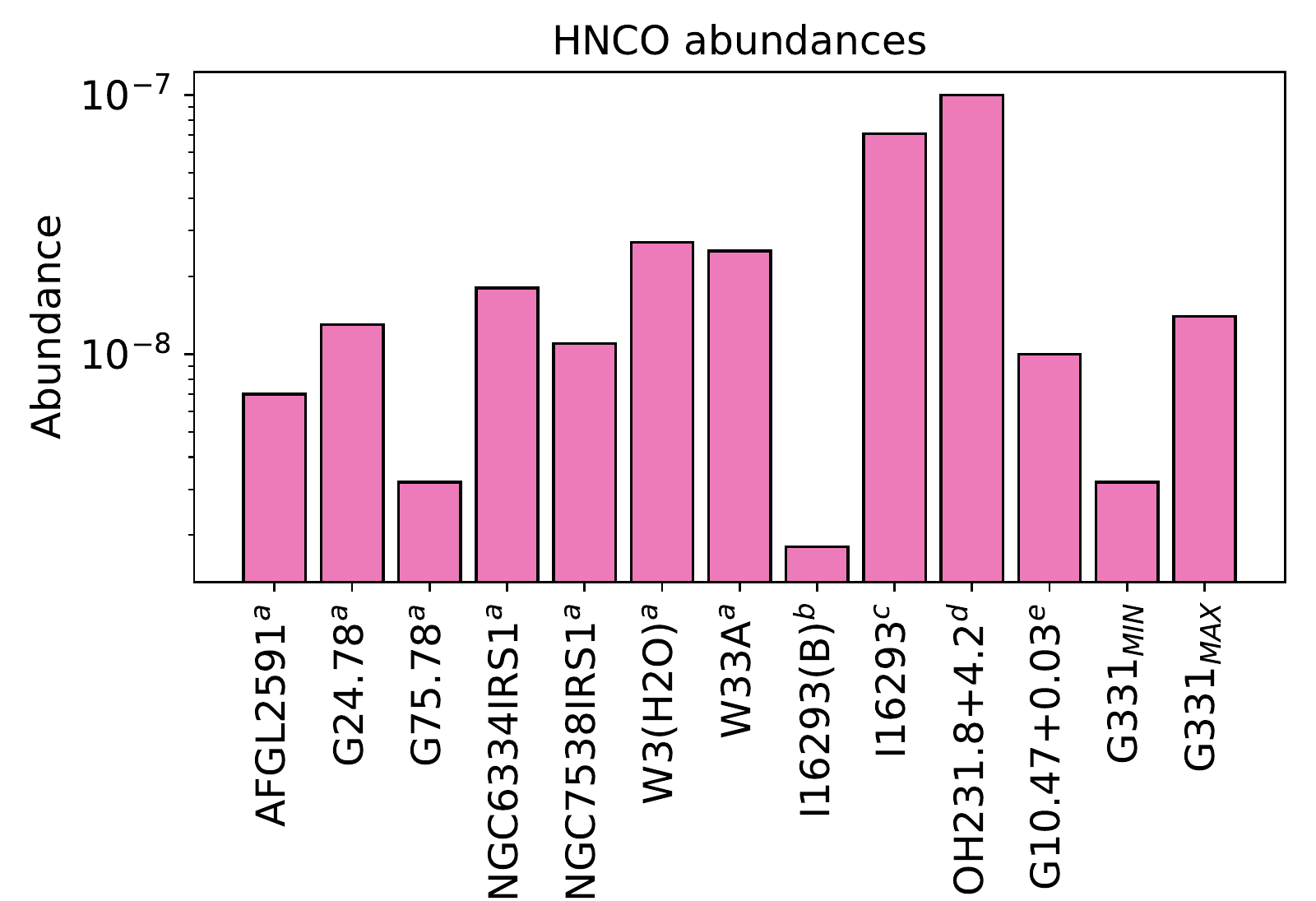}
    \caption{Histograms of the column densities of HNCO (top) and their relative abundances to H$_2$  (bottom) in several sources: $^a$\citet{Bisschop07},  $^b$IRAS 16293--2422 \citep[B, ][]{Martin17}, $^c$IRAS 16293--2422 \citep[\textit{hot corino}, ][]{Hernandez19}, $^d$\citet{Prieto15}, $^e$\citet{Gorai2020}. The minimum and maximum values (G331$_{MIN}$ and G331$_{MAX}$)  obtained for G331 are also shown.} 
    \label{fig:hist-ab}
\end{figure}

In Fig.~\ref{fig:hist-ab}, the histograms of the column densities and HNCO abundances for several sources are displayed in order to compare with those derived for G331. The objects shown are high-mass young stellar objects (YSOs) in hot core stages \citep[AFGL 2591, G24.78, G75.78, NGC 6334 IRS1, NGC 7538 IRS1, W3(H2O) and W 33A,][]{Bisschop07}; low mass protostar IRAS 16293--2422, in the compact source B \citep{Martin17} and in the hot corino region  \citep{Hernandez19}; oxygen-rich circumstellar envelope around an intermediate-mass evolved star \citep[OH231.8+4.2,][]{Prieto15}; and hot molecular core \citep[G10.47+0.03,][]{Gorai2020}.

Concerning the HNCO column densities, G331 and most of the YSOs of \citet{Bisschop07}  present similar values. This might be due to similar evolutionary stage of these objects, with  protostars embedded in hot molecular cores. The two sources of IRAS 16293--2422 have higher column densities, which could be connected to the innermost and densest regions of this object (e.g. hot corino). In colder envelopes, the HNCO column densities could be up to four orders of magnitude lower \citep{Hernandez19}. The histogram of the abundances, on the other hand, reveals  a greater variation among the objects. 
The HNCO abundances of G75.78 and G331$_{MIN}$ are very similar, while G331$_{MAX}$ is in better agreement with the other YSOs.

It is worth noting that the column density reported for G10.47+0.03 is much higher than G331, however the abundance of HNCO in this source is in the range calculated for G331. The most evolved object in this sample, OH231.8+4.2, shows  the lowest column density and the highest abundance of HNCO. Despite these peculiar cases, the values of HNCO  column density and  abundance in G331 are in agreement with  those obtained in most of the hot molecular cores  in the histograms.

\subsection{The CHNO isomers: HNCO, HOCN, HCNO and HONC} \label{sec:4.2}

HNCO has three  meta-stable isomers of which HCNO (fulminic acid) and HOCN (cyanic acid) as well as HNCO have been observed in the Interstellar Medium. Isofulminic acid (HONC) is the only meta-stable isomer that has not been detected in the ISM yet.  All these isomers are quasi-linear molecules except isofulminic acid, which has a bent structure~\citep{Mladenovic2008,Mladenovic2009}. HCNO and HOCN are much less abundant than HNCO and their occurrence in space has been restricted to a few sources. On the one hand, HCNO was discovered in dark clouds by \citet{Marcelino2009}; using observations at the 3~mm band of  the IRAM-30m telescope, they detected HCNO in the starless cores B1, L1544, L183 as well as in the low mass star forming region L1527. On the other hand, HOCN was observed towards the Galactic centre by \citet{Brunken2010}; similarly, using observations at the 3~mm band of IRAM-30m, they detected transitions of HOCN in a quiescent molecular cloud and in several positions of the Sgr~B2 complex.

The dipole moments of the four  isomers have been reported and discussed in the literature. The values for HNCO, HCNO,  HOCN and HONC have been estimated to be $\sim$~2.08 , 3.1,  3.7 and 3.13~D, respectively (e.g. \citealt{Hocking1974,Hocking1975,Takashi1989,Mladenovic2008,Mladenovic2009}). Among the four species, HNCO is the most stable one (by 1.1~eV), this has been demonstrated through studies about the equilibrium structure and energetic of the CHNO isomers (e.g. \citealt{Mladenovic2008,Mladenovic2009}). Regarding the abundance ratios between the isomers, \citet{Marcelino2009} determined HNCO/HCNO between 40 -- 70 in dark clouds and starless cores;   \citet{Brunken2010} determined values of $\sim$~0.3 -- 0.8~per~cent for HOCN relative to HNCO towards the Galactic centre.

In this work, HNCO was  clearly detected through several spectral lines, as  presented in Table~\ref{tab:lines-v2}. The search for HCNO and HOCN, on the other hand, did not yield conclusive results to confirm their presence in G331. About HOCN, the number of predicted transitions with $E_u <$ 450~K is about 60; however, only spectral noise has been  observed where most of the lines should appear. Two transitions might be speculatively  inspected  since they are dominated by spectral emission relatively well known in the literature. That is the case of the transition HOCN 9$_{3,6}$--8$_{3,5}$ at the rest frequency 188655.358~MHz, which is likely affected by SO$_2$ 9$_{2,8}$--9$_{1,9}$ at 188654.973~MHz; and HOCN 12$_{3,10}$--11$_{3,9}$ at the rest frequency 	251528.269~MHz, which is likely affected by c-C$_3$H$_2$ 6$_{2,5}$--5$_{1,4}$ at 251527.311~MHz \citep{Nummelin1998}. Regarding HCNO, the number of predicted transitions is considerably minor in comparison to HOCN,  about 10 for the same interval of $E_u$. Similarly, the transition HCNO 12--11 is identified at the rest frequency 275227.774~MHz, however it is likely affected by SO$_2$ 15$_{3,13}$--15$_{2,14}$ at 275240.184
~MHz \citep{Lor84}.  Additionally, in the SO$_2$ case, a factor that also makes difficult a positive identification is that the  lines exhibit  Lorentzian profiles, so that such emission could be more related to the outflow. In the c-C$_3$H$_2$ case, the line exhibit  a Gaussian profile as the HNCO lines, which could be more in agreement with the physical origin of the HNCO emission. In perspective, the \ \lq\lq puzzle\rq\rq \ of the CHNO isomers in hot molecular cores is compelling, theoretical and/or experimental approaches might be required for a better comprehension of their properties in the ISM.

\subsection{Internal partition function of CHNO isomers}
\label{subsec:Qrv}

The rotational-vibrational partition function of the isomers of CHNO (\S~\ref{sec:4.2})~\citep{McLean1977} have been updated in this work according to \citet{Carvajal2019} and presented here altogether for their comparison.  

Usually, the internal partition function is computed as a direct sum of a considerable number of rovibrational energies in order to reach a convergence at any temperature. Unfortunately, the spectroscopic data for the isomers of CHNO is still insufficient. Then, we have to compute  the partition function adopting some suitable approximations that were already validated \citep[see, e.g.,][and references therein]{Carvajal2019}. Thus,
the molecular partition functions for CHNO isomers are computed as a product of the rotational partition function $Q_{\rm rot}(T)$ and of the vibrational contribution $Q_{\rm vib}(T)$:

 \begin{equation}
\label{eq:rtvQapprox}
Q_{\rm rv}(T) \approx Q_{\rm rot}(T) \, Q_{\rm vib}(T) ~~,
\end{equation}

\noindent where the rotational contribution is computed as a direct sum whether the convergence is reached with the available spectroscopic data. In the particular cases of the CHNO isomers, the rotational spectroscopic information is available in the CDMS catalogue and the predicted rotational energies in the vibrational ground state can be computed with Pickett's code~\citep{Pickett1991}. 
Otherwise, the approximation of the classical partition function can be a suitable alternative to provide the rotational partition function for sources at not very low temperatures~\citep{Herzberg} besides being a useful tool to test the convergence of the direct sum at higher temperatures~\citep{Carvajal2019}. In the particular cases of CHNO isomers, the classical partition function values turn out to be adequate at temperatures higher than 20~K. However, the addition of some simple correction terms to the classical partition function can improve this approximation considerably~\citep{Wells2020}. For all the CHNO isomers, the nuclear spin statistical weights are the same and, in this work, they are considered as 1.

\begin{table*}
\caption{Vibrational and rotational-vibrational partition function for CHNO isomers: isocyanic acid (HNCO); cyanic acid (HOCN); fulminic acid (HCNO); and isofulminic acid (HONC). Comparison between the values
  obtained in the present study and those published in CDMS catalogue$^a$.}
\label{tab-PartF}
\begin{center}
{\footnotesize
\begin{tabular}{c|cccc|cccc}
\hline\hline
&   \multicolumn{4}{c}{HNCO$^b$} &   \multicolumn{4}{c}{HOCN}  \\ 
\hline
$T$(K) &$Q_{\rm vib}^{\rm harm}$~$^c$ & $Q_{\rm rv}$\citep{Carvajal2019}$^d$ & $Q$(CDMS)$^e$ & Rel. Diff.(\%)$^f$ &$Q_{\rm vib}^{\rm harm}$~$^c$ & $Q_{\rm rv}$(Present work)$^{d,g}$ & $Q$(CDMS)$^e$ & Rel. Diff.(\%)$^f$ \\ \hline
  2.725&1.000000  &    5.5129(50)   &   5.51 & 0.00 &  1.000000 &     5.7601(52) &   5.7601  & 0.00 \\
  5.000  &1.000000  &    9.8228(42)   &   9.82 & 0.00 &  1.000000 &    10.3038(44) &  10.3038 & 0.00  \\
  9.375&1.000000  &   18.4493(36)   &  18.45 & 0.00 &  1.000000 &    20.1244(37) &  20.1244 & 0.00  \\
 18.750 &1.000000  &    42.8295(30)   &  42.83 & 0.00 &  1.000000 &    50.8161(32) &  50.8159 & 0.00 \\
 37.500 &1.000000  &  117.3053(27)   & 117.30 & 0.01 &  1.000000 &   142.1565(12) & 142.1413  &0.01  \\
 75.000  &1.000019  &  332.0002(27)   & 331.99 & 0.00 &  1.000314 &   402.21(33)  & 401.3740  & 0.21 \\
150.000  &1.006395  &  949.759(40)   & 943.71 &  0.64 &  1.024660 &  1165.30(98) &  1135.3539 & 2.57 \\ 
225.000  &1.048753  & 1827.01(36)   &1742.43 &  4.63 &  1.114711 &  2328.9(22) &  2087.4819 & 10.37 \\
300.000  &1.144368  & 3080.9(12)   &2695.34 & 12.51 &  1.269376 &  4083.1(48) &  3217.0765 & 21.21 \\
500.000  &1.667813  & 9690.5(89)   &5866.52 & 39.46 &  1.984226 & 13733.(23) &  ----- & ----- \\
\hline
\hline
&   \multicolumn{4}{c}{HCNO} &   \multicolumn{4}{c}{HONC}  \\ 
\hline
$T$(K) &$Q_{\rm vib}^{\rm harm}$~$^c$ & $Q_{\rm rv}$(Present work)$^d$ & $Q$(CDMS)$^e$ & Rel. Diff.(\%)$^f$ &$Q_{\rm vib}^{\rm harm}$~$^c$ & $Q_{\rm rv}$(Present work)$^{d,h}$ & $Q$(CDMS)$^e$ & Rel. Diff.(\%)$^f$ \\ \hline
    2.725 &    1.000000  &     5.2981(49) & 5.2980 & 0.00 &   1.000000 &      5.5525(50) & 5.5307 & 0.39 \\
    5.000 &    1.000000  &     9.4247(41) & 9.4247 &  0.00 &   1.000000 &      9.9416(42) & 9.9014 & 0.40 \\
    9.375 &    1.000000  &    17.3697(34) & 17.3697 & 0.00 &   1.000000 &     19.7403(36) & 19.6616 & 0.40 \\
   18.750 &    1.000000  &    34.4006(29) & 34.4005 & 0.00 &   1.000000 &     50.9055(31) & 50.7102 & 0.38 \\
   37.500 &    1.000369  &    68.4919(25) & 68.4665 & 0.04 &   1.000033 &    142.7292(30) & 142.2148 & 0.36 \\
   75.000 &    1.027795  &   140.402(52) & 136.6050 & 2.70 &   1.007476 &    405.760(45) & 401.5499 & 1.04 \\
  150.000 &    1.296184  &   353.73(63) & 272.9036 & 22.85 &   1.130785 &   1286.6(11) & 1135.8099 & 11.72 \\
  225.000 &    1.842269  &   753.9(21) & 409.2300 & 45.72 &   1.393281 &   2911.3(48) & 2088.3400 & 28.27 \\
  300.000 &    2.707133  &  1477.0(53) & 545.5841 & 63.06 &   1.764826 &   5676.3(125)& 3218.5016 & 43.30 \\ 
  500.000 &    7.363415  &  6695(31) &  ----- & ----- &   3.328126 &  23019.(129)&  ----- & ----- \\
\hline\hline
\end{tabular}

\begin{flushleft}
$^a$ The  nuclear spin degeneracy is considered as $1$.

$^b$ Result from \cite{Carvajal2019}.

$^c$ The vibrational partition function is  computed with the harmonic approximation. For more details, see the text.

$^d$ $Q_{\rm rv}$=$Q_{\rm rot}(\mbox{\rm direct sum}) \, Q_{\rm
    vib}^{\rm harm}$ unless noted otherwise. An upward estimate of the uncertainties is given in parentheses in  units of the last quoted digits. For more details, see the text. 

$^e$ Rotational partition function computed as a direct sum with no vibrational contribution. Their values are reported in CDMS catalogue~\citep{end2016}.

$^f$ Relative difference of the partition function given in the present study with respect to the one reported in CDMS catalogue.

$^g$ The rotational contribution of the partition function is computed with a direct sum up to T~=~18.75~K inclusive. From 37.50~K forward, the rotational partition function is obtained with the classical approximation. For more details, see the text.

$^h$ The rotational contribution of the partition function is computed with a direct sum except for T~=~500~K, which is obtained with the classical approximation. For more details, see the text.

\end{flushleft}
}
\end{center}
\end{table*}

Concerning the vibrational partition functions for the CHNO isomers, these cannot be obtained as a direct sum because there are not enough experimental or calculated vibrational energies at disposal. Therefore, the harmonic approximation~\citep{Herzberg} 
 has been used to estimate the vibrational contribution of the partition function and, therefore, it is enough to know the vibrational fundamental frequencies. 
If any bending degree of freedom for these four isomers was susceptible to undergo a strong anharmonicity caused, e.g., by a large amplitude motion, the vibrational partition function could be split into two contributions \citep[see, e.g.,][and references therein]{Carvajal2019}: the one derived from small amplitude vibrations and another coming from the large amplitude vibrational modes. The vibrational partition function from the small amplitude vibrations can be computed with the harmonic approximation whereas the large amplitude contribution should be calculated as a direct sum using the predicted energies of the large amplitude bending states~\citep{Favre2014}, which could be calculated with any approach conceived for the analysis of nonrigid molecules \citep[see, e.g.,][]{Jensen1983,KhaloufRivera2020}.

Table~\ref{tab-PartF} provides, for the four CHNO isomers, the results of the vibrational contribution in the harmonic approximation ($Q_{\rm vib}^{\rm harm}$) of the internal partition function ($Q_{\rm rv}$), the internal partition function  computed in the present work by means of Eq.~(\ref{eq:rtvQapprox}), the rotational partition function in the CDMS catalogue computed as a direct sum~\citep{end2016} and the relative differences between the values reported in the present work and in the CDMS for  the temperature range from 2.725~K to 500~K. In addition, the uncertainties of the internal partition functions have  been included for the four CHNO isomers. These have been estimated upwardly considering large uncertainties for the rotational and vibrational fundamental energies at 100~MHz and 1~cm$^{-1}$, respectively. In general, at temperatures above 20~K, the relative difference between the partition function values obtained in the present work and those in CDMS catalogue are substantially larger than the upward estimate of the uncertainty of the internal partition function. 

From Table~\ref{tab-PartF}, it can be observed that the rotational contributions to the internal partition functions of the four isomers have similar values at the lowest temperatures, e.g. down to $T$~=~10~K. For higher temperatures,  above $T$~=~225~K, the difference between the rotational partition function of fulminic acid and the values of the other three isomers (isocyanic acid, cyanic acid and isofulminic acid), with  more similar values, becomes larger. This could be explained because of the structure of fulminic acid is linear and of the other have a  near-prolate asymmetric top structure. Nevertheless, the internal partition functions of isocyanic acid, cyanic acid and isofulminic acid become completely different at higher temperatures despite the fact that these three isomers have a bent structure. This stems from the differences among the vibrational partition functions of the bent isomers.

In Table~\ref{tab-PartF}, the weight of the vibrational contribution to the internal partition function can also be assessed for the set of the four CHNO isomers at different temperatures. The differences between the internal partition function and the rotational partition function become noteworthy (more than 1~per~cent) at temperatures higher than 225~K for HNCO, 150~K for HOCN, and  $T$~=~75~K for HCNO and HONC.
Even when the vibrational partition function has been calculated with the harmonic approximation and, in some cases, using  fundamentals, the values of this approximation, although underestimated, are in good agreement with the direct sum values~\citep{Carvajal2019}.

As supplementary material, the rotational, vibrational and rovibrational partition function calculated in the present work from 1~K to 500~K (in intervals of 1~K) is also provided for the following isomers: cyanic acid (HOCN); fulminic acid (HCNO); and isofulminic acid (HONC). For isocyanic acid (HNCO), same material can be found elsewhere~\citep{Carvajal2019}.

In App.~\ref{appendix-Qrv} we will give some details about the calculation of the partition function for the four CHNO isomers.

\subsection{Chemical modelling with Nautilus}

In order to better understand the chemical pathways involved in the production and destruction of HNCO, we used the \textsc{Nautilus} code \citep{nautilus} to carry out a time-dependent chemical model. \textsc{Nautilus} allows to simulate and study grain- and gas-phase chemical processes for instance in hot and cold cores \citep{sem10,reb14,rua15,nautilus}. The predictions of the model include the time evolution of the chemical abundances for a given set of physical and chemical parameters. For the solid-state chemistry, it considers mantle and surface as chemically active, following the formalism of \citet{has93} and the experimental results of \citet{ghe15}. In this sense, the code can perform a three-phase (gas plus grain mantle and surface) time-dependant simulation of the chemistry in hot cores, including chemical reactions in both gas and solid phases \citep{rua15}. 

This approach is interesting for our work since most of HNCO may be trapped in icy grains, as suggested by \citet{Altwegg20}. The grain chemistry of the code considers the standard direct photo-dissociation by photons along with the photo-dissociation induced by secondary UV photons, as described by \citet{Prasad83}, which are effective processes on the surface and mantle of the grains. The network of chemical reactions used by the code is presented on the KIDA\footnote{\url{http://kida.obs.u-bordeaux1.fr/}} (KInetic Database for Astrochemistry) catalogue \citep{Wak15}.

\begin{table}
	\caption{Initial abundances assumed for the \textsc{Nautilus} model (extracted from \citet{vid18}). References are: 1. \citet{wak08}; 2. \citet{Jenkins09}; 3. \citet{Hincelin11}; 4. low-metal abundances from \citet{Graedel82}; 5. depleted value from \citet{Neufeld05}.}
	\label{tab:abs}
    \centering
	\begin{tabular}{lcclcc}
		\hline
		Element & {n$_{i}$/n$_{H}$}$^{a}$ & Ref. & Element & {n$_{i}$/n$_{H}$}$^{a}$ & Ref. \\
		\hline
		H$_2$ & 0.5 & &  He & 9.0(-2) & 1 \\
		N  & 6.2(-5) & 2 &  O  & 2.4(-4) & 3 \\
		C$^+$ & 1.7(-4) & 2 & S$^+$ & 1.5(-5) & 2 \\
		Fe$^+$ & 3.0(-9) & 4 & Si$^+$ & 8.0(-9) & 4 \\
		Na$^+$ & 2.0(-9) & 4 & Mg$^+$ & 7.0(-9) & 4 \\
		Cl$^+$ & 1.0(-9) & 4 & P$^+$ & 2.0(-10) & 4 \\
        F & 6.7(-9) & 5 & & & \\
		\hline
		\multicolumn{6}{l}{$^a$ Abundances given in the format a(b) representing a$\times$10$^{b}$. }
	\end{tabular}
\end{table}

The first and simplest simulation to perform is a zero-dimensional, i.e., total density, gas temperature, and other physical condition are uniform within the considered core and throughout the simulation time. There is no structure evolution in this case.  Once our source present a shell-like structure, a more elaborated structure model is necessary to be simulated together with \textsc{Nautilus} in order to represent the HNCO abundances more accurately. Nevertheless, this approximation can furnish the first insights in the chemistry of G331.  The initial elemental abundances for the core were selected from the recent values used by \citet{vid18} and are displayed in Table~\ref{tab:abs}. They represent an intermediate-metal abundance case, in which all the abundances are the same as the low-metal one except that the amount of the element sulphur is raised to 1.5$\times$10$^{-5}$ compared with H$_2$ \citep{wak06}. Concerning parameters such as the visual extinction, we assumed a typical value of 10~mag for a dark cloud. The gas and dust temperatures were considered the same in the simulations.

\begin{figure}
\includegraphics[width=\columnwidth,keepaspectratio]{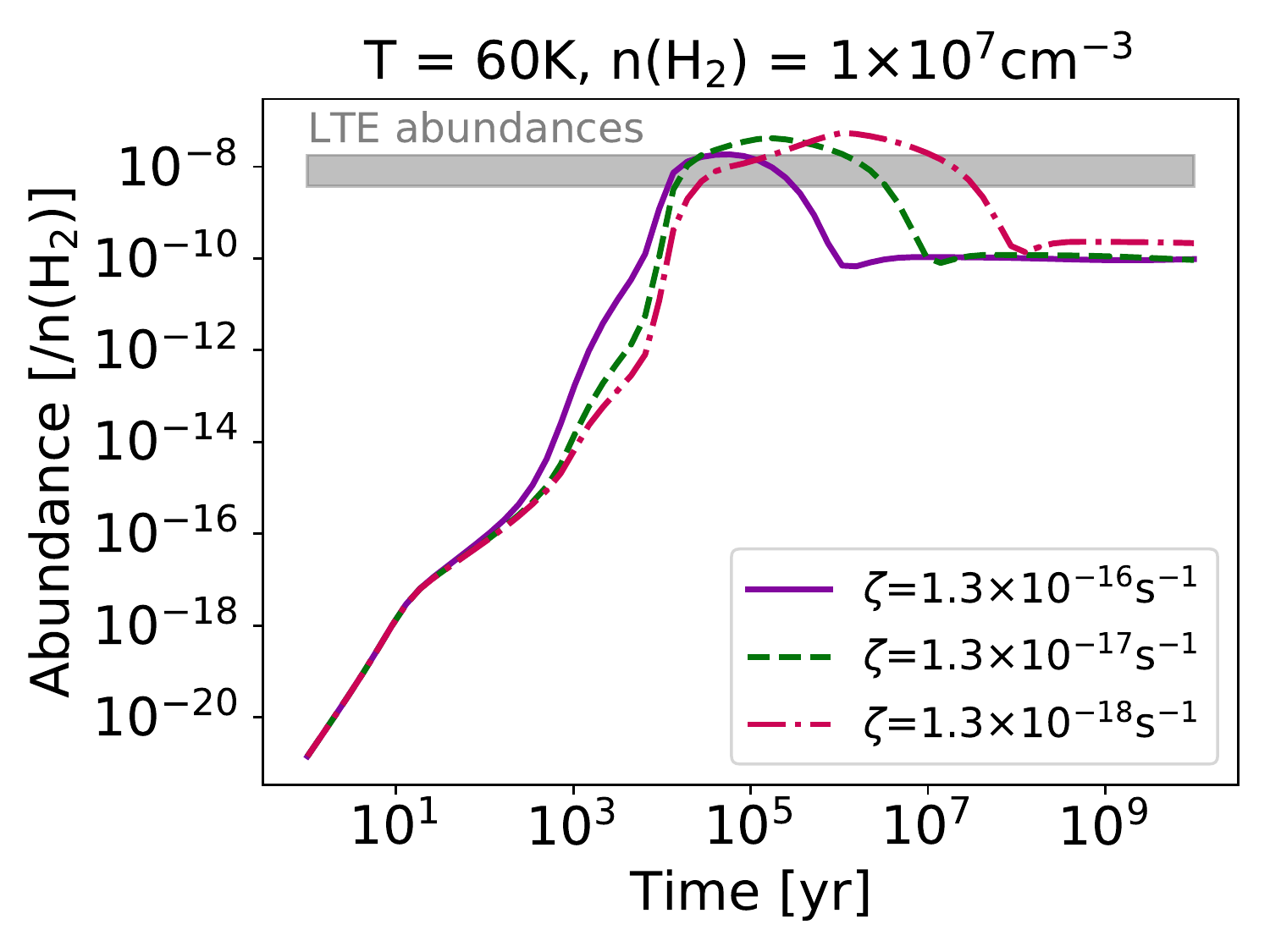}
\caption{Best-fit time evolution of the HNCO abundances simulated with \textsc{Nautilus}. We assumed a gas and grain temperatures of 60~K, and density of 10$^7$~cm$^{-3}$. The cosmic ray ionisation rate ($\zeta$) was varied in 1.3$\times$10$^{-16}$, 1.3$\times$10$^{-17}$ and 1.3$\times$10$^{-18}$s$^{-1}$. The grey strip indicates the range of the derived LTE abundances.}
\label{fig:naut-final}
\end{figure}

Several simulations were tested for a range of density and temperature of 10$^6$~--~10$^8$~~cm$^{-3}$ and 50~--~200~K, respectively \citep[see][for more details]{thesis-canelo}. The best-fit model is obtained adopting  a density of 1$\times$10$^7$~cm$^{-3}$ and a temperature of 60~K, and is presented in Fig.~\ref{fig:naut-final}.  The standard cosmic ray ionisation rate for H$_2$ ($\zeta_{\text{H$_2$}}$) is 1.3$\times$10$^{-17}$s$^{-1}$ \citep{Wak15,nautilus}, which is a key parameter for the gas-grain models since cosmic rays induce desorption processes in which chemical species pass to the gas phase from grain surfaces \citep{has93,rua15}. Taking into account that the cosmic ray ionisation might vary in interstellar gas clouds, e.g. 0.6--6 $\times$ 10$^{-17}$~s$^{-1}$ \citep{vandertak2000,Dalgarno2006}, we also tested models considering $\zeta_{\text{H$_2$}}$ = 1.3$\times$10$^{-16}$ and 1.3$\times$10$^{-18}$s$^{-1}$, the results are shown in Figure~\ref{fig:naut-final}. The main difference between the models is the plateau of the abundance intensity peaks and, consequently, the position of the maximum value. The maximum abundances obtained were 1.86, 4.19 and 5.48 $\times$10$^{-8}$, respectively. Although these abundances are a bit higher than those obtained from the rotational diagrams, the model at 60~K is the only one that can reproduce the abundance range of G331 and also the excitation temperature derived for HNCO, although the final adopted density is slightly higher than the core density derived by \citet[][see Sec.~\ref{intro:G331}]{Hervias19}.    

From Fig.~\ref{fig:naut-final} it can be seen that the balance of the chemical reaction and chemical equilibrium are highly dependant on the cosmic ionisation rate. Higher rates induce more chemical reactions (as expected, as it is a more significant energy source) that last a shorter period of time. Maximum abundances are also lower but the chemical equilibrium occurs faster. Considering $\zeta$ = 1.3 $\times$10$^{-17}$s$^{-1}$, the chemical age could be between 3 $\times$10$^{4-6}$ years with a maximum simulated abundance around 10$^{5}$ years. The other limiting values of $\zeta$ presented the same abundance 1.4 $\times$10$^{-8}$ derived for G331 at 10$^{5}$ years, approximately. These models could suggest that 10$^{5}$ years may be a plausible chemical age for G331. Hot cores  and hot corinos, in general, are expected to have a dynamical age up to 10$^{4-5}$~years, as obtained for IRAS 16293-2422 \citep{Hernandez19} and G10.47+0.03 \citep{Gorai2020}, for example. Within this scenario, the chemical age derived for G331 is also in agreement with most of the objects of Fig.~\ref{fig:hist-ab}. The molecular flow of the evolved star OH231.8+4.2 has an estimated dynamical age of 800~yr, but this source is in a later evolutionary stage. Nevertheless, we cannot conclude that HNCO could be used as a chemical clock. Our simulations are zero-dimensional and this molecule can be highly dependent on the initial parameters \citep{Hernandez19}. Moreover, the chemical age obtained with \textsc{Nautilus} is, at least, one order of magnitude  larger than the estimated dynamical age of G331 \citep[$\sim$~2000 years, ][]{Merello13a}. 

In these simulations, the code uses two formation pathways which lead to HNCO by means of gas-grain processes. The whole mechanism suggest that HNCO is  initially formed on the surface of grains, via H $+$ OCN $\rightarrow$ HNCO, then it might pass from the solid (s) to the gaseous (g) phase through sublimation processes, via HNCO (s) $\rightarrow$ HNCO (g). They are enough to reproduce the observed abundances in this temperature regime and with a zero-dimension simulation, but the code returns a greater time scale to do it so. Furthermore, higher temperatures presented a drastic decrease in the simulated abundance.  For a more robust model,  with structure evolution and time evolution of physical conditions, other reactions may have to be included in the code, if their rate coefficients are available. One example is the reaction NH $+$ CO $\rightarrow$ HNCO, normally proposed as the main pathway to form HNCO in analogous mixtures of interstellar ice processed by proton or UV radiation \citep{Fedoseev15}. It is also possible that the incorporation of this reaction in the code may lead to a chemical age similar to the previously derived dynamical age of G331. 


\section{Conclusions}
\label{sec:conc}

In this paper, we reported the observation of 42 emission lines of the potential prebiotic molecule HNCO in the hot molecular core G331. This core was observed with the APEX telescope covering the interval frequency 160 -- 355~GHz. The HNCO transitions are distributed into their $K_a$-ladder numbers of 0, 1 and 2. The $K_a$~=~0 transitions presented the highest fluxes while the $K_a$~=~2 transitions showed the lowest ones, appearing as spectral pairs partially resolved. Such trends are expected from the different $K_a$-ladder numbers, in which the spectral pairs of each ladder get a better separation as the frequency increase. Furthermore, the observations allowed to see  a particular spectral profile with extended tails in some HNCO lines, as in the HNCO  $J$~=~8-7 and $J$~=~15-14 transitions. The HNCO emission can be responsible for about 80~per~cent of the entire flux, while the wing could be an effect of the outflow or a contaminant emission.

With LTE rotational diagrams, we obtained the HNCO excitation temperature and column density of $T_{ex}$= 59.4 $\pm$ 2.3~K and $N$(HNCO) = (3.1 $\pm$ 0.4) $\times$ 10$^{15}$~cm$^{-2}$, without the opacity correction, and  $T_{ex}$= 58.8 $\pm$ 2.7~K and  $N$(HNCO) =  (3.7 $\pm$ 0.5) $\times$ 10$^{15}$~cm$^{-2}$,  considering beam dilution effects. This value of temperature could indicate that HNCO molecules are located in the external and colder regions of G331, according to the shell-like structure model of G331. We also derived the  HNCO relative abundances by considering two main values, so that it was obtained values in the interval of (3.8 $\pm$ 0.5) $\times$ 10$^{-9}$ < [HNCO] <  (1.4 $\pm$ 0.2) $\times$ 10$^{-8}$. We compared our column densities and abundances of HNCO with values from other hot molecular cores and evolved objects, and our results are in agreement with most of the objects.

In addition, the internal partition functions of the four CHNO isomers (isocyanic acid, HNCO; cyanic acid, HOCN; fulminic acid, HCNO; and isofulminic acid, HONC) have been updated with the vibrational contribution and their values are provided from 1~K to 500~K in intervals of 1~K in the supplementary material. We think that this update is relevant for the estimate of abundances of the four CHNO isomers in surveys at temperatures higher than 225~K for HNCO, 150~K for HOCN, and  $T$~=~75~K for HCNO and HONC.

Finally, we simulated the chemistry of HNCO with the three-phase time-dependent code  \textsc{Nautilus}. We used a zero-dimensional simulation with a temperature of 60~K that represented our derived abundances without the need to add more grain-phase reactions to the  \textsc{Nautilus} database. We also varied the cosmic ionisation rate to better comprehend the influence of this physical parameter in the simulations. From the models, we could also suggest a chemical age around 10$^5$ years for G331, which is higher than estimated dynamical age. One explanation for this difference could be the absence of a key reaction to form HNCO in the code. 

\section*{Acknowledgements}


The authors thank the anonymous referee for the useful comments that improved the article. C.M.C. acknowledges the support of CNPq, Conselho Nacional de Desenvolvimento Cient\'ifico e Tecnol\'ogico - Brazil, process number 141714/2016-6. This study was financed in part by the Coordena\c{c}\~ao de Aperfei\c{c}oamento de Pessoal de N\'ivel Superior - Brasil (CAPES) - Finance Code 001. L. B.  acknowledges support from CONICYT project Basal AFB-170002. EM acknowledges support from the Brazilian agencies FAPESP (grant 2014/22095-6) and CNPq (grant 150465/2019-0).
M.C. acknowledges the financial support from the European Union's
Horizon 2020 research and innovation program under the Marie Sk\l odowska-Curie
grant agreement No 872081; from the Spanish National Research, Development,
and Innovation plan (RDI plan) under the project PID2019-104002GB-C21; the Consejer\'{\i}a de Conocimiento, Investigaci\'on y Universidad, Junta de Andaluc\'{\i}a and European Regional Development Fund (ERDF), ref.  SOMM17/6105/UGR; the Ministerio de Ciencia,
Innovaci\'on y Universidades (ref.COOPB20364); and by the Centro de Estudios
Avanzados en F\'{\i}sica, Matem\'aticas y Computaci\'on (CEAFMC) of the
University of Huelva.


\section*{Data Availability}

The data underlying this article will be shared on reasonable request to the corresponding author.



\bibliographystyle{mnras}
\bibliography{canelo-hnco-2021} 

\begin{thebibliography}{}
\makeatletter
\relax
\def\mn@urlcharsother{\let\do\@makeother \do\$\do\&\do\#\do\^\do\_\do\%\do\~}
\def\mn@doi{\begingroup\mn@urlcharsother \@ifnextchar [ {\mn@doi@}
  {\mn@doi@[]}}
\def\mn@doi@[#1]#2{\def\@tempa{#1}\ifx\@tempa\@empty \href
  {http://dx.doi.org/#2} {doi:#2}\else \href {http://dx.doi.org/#2} {#1}\fi
  \endgroup}
\def\mn@eprint#1#2{\mn@eprint@#1:#2::\@nil}
\def\mn@eprint@arXiv#1{\href {http://arxiv.org/abs/#1} {{\tt arXiv:#1}}}
\def\mn@eprint@dblp#1{\href {http://dblp.uni-trier.de/rec/bibtex/#1.xml}
  {dblp:#1}}
\def\mn@eprint@#1:#2:#3:#4\@nil{\def\@tempa {#1}\def\@tempb {#2}\def\@tempc
  {#3}\ifx \@tempc \@empty \let \@tempc \@tempb \let \@tempb \@tempa \fi \ifx
  \@tempb \@empty \def\@tempb {arXiv}\fi \@ifundefined
  {mn@eprint@\@tempb}{\@tempb:\@tempc}{\expandafter \expandafter \csname
  mn@eprint@\@tempb\endcsname \expandafter{\@tempc}}}

\bibitem[\protect\citeauthoryear{Albert, Winnewisser  \& Winnewisser}{Albert
  et~al.}{1996}]{Albert1996}
Albert S.,  Winnewisser M.,   Winnewisser B.,  1996, Ber. Bunsenges. Phys.
  Chem., 100, 1876

\bibitem[\protect\citeauthoryear{{Altwegg} et~al.,}{{Altwegg}
  et~al.}{2020}]{Altwegg20}
{Altwegg} K.,  et~al., 2020, \mn@doi [Nature Astronomy]
  {10.1038/s41550-019-0991-9}, \href
  {https://ui.adsabs.harvard.edu/abs/2020NatAs.tmp....3A} {p.~3}

\bibitem[\protect\citeauthoryear{{Armstrong} \& {Loren}}{{Armstrong} \&
  {Loren}}{1984}]{Arm84}
{Armstrong} T.,  {Loren} R.~B.,  1984, Tech. Rep. AST, 8116403-1

\bibitem[\protect\citeauthoryear{{Bally}}{{Bally}}{2016}]{Bally2016}
{Bally} J.,  2016, \mn@doi [\araa] {10.1146/annurev-astro-081915-023341}, \href
  {https://ui.adsabs.harvard.edu/abs/2016ARA&A..54..491B} {54, 491}

\bibitem[\protect\citeauthoryear{{Belitsky} et~al.,}{{Belitsky}
  et~al.}{2018}]{belitsky2018}
{Belitsky} V.,  et~al., 2018, \mn@doi [\aap] {10.1051/0004-6361/201731458},
  \href {https://ui.adsabs.harvard.edu/abs/2018A&A...612A..23B} {612, A23}

\bibitem[\protect\citeauthoryear{{Bisschop}, {J{\o}rgensen}, {van Dishoeck}  \&
  {de Wachter}}{{Bisschop} et~al.}{2007}]{Bisschop07}
{Bisschop} S.~E.,  {J{\o}rgensen} J.~K.,  {van Dishoeck} E.~F.,   {de Wachter}
  E.~B.~M.,  2007, \mn@doi [\aap] {10.1051/0004-6361:20065963}, \href
  {https://ui.adsabs.harvard.edu/abs/2007A&A...465..913B} {465, 913}

\bibitem[\protect\citeauthoryear{{Biver} et~al.,}{{Biver}
  et~al.}{2006}]{Biver06}
{Biver} N.,  et~al., 2006, \mn@doi [\aap] {10.1051/0004-6361:20053849}, \href
  {https://ui.adsabs.harvard.edu/abs/2006A&A...449.1255B} {449, 1255}

\bibitem[\protect\citeauthoryear{{Bronfman}, {Garay}, {Merello}, {Mardones},
  {May}, {Brooks}, {Nyman}  \& {G{\"u}sten}}{{Bronfman}
  et~al.}{2008}]{Bronfman08}
{Bronfman} L.,  {Garay} G.,  {Merello} M.,  {Mardones} D.,  {May} J.,  {Brooks}
  K.~J.,  {Nyman} L.-{\r{A}}.,   {G{\"u}sten} R.,  2008, \mn@doi [\apj]
  {10.1086/522487}, \href
  {https://ui.adsabs.harvard.edu/abs/2008ApJ...672..391B} {672, 391}

\bibitem[\protect\citeauthoryear{Brunken, Gottlieb, McCarthy  \&
  Thaddeus}{Brunken et~al.}{2009}]{Brunken2009}
Brunken S.,  Gottlieb C.,  McCarthy M.,   Thaddeus P.,  2009, \apj, 697, 880

\bibitem[\protect\citeauthoryear{{Br{\"u}nken}, {Belloche}, {Mart{\'\i}n},
  {Verheyen}  \& {Menten}}{{Br{\"u}nken} et~al.}{2010}]{Brunken2010}
{Br{\"u}nken} S.,  {Belloche} A.,  {Mart{\'\i}n} S.,  {Verheyen} L.,   {Menten}
  K.~M.,  2010, \mn@doi [\aap] {10.1051/0004-6361/200912456}, \href
  {https://ui.adsabs.harvard.edu/abs/2010A&A...516A.109B} {516, A109}

\bibitem[\protect\citeauthoryear{Canelo}{Canelo}{2020}]{thesis-canelo}
Canelo C.~M.,  2020, PhD thesis, IAG-USP, Brasil,
  \mn@doi{https://doi.org/10.11606/T.14.2020.tde-03022021-131439}

\bibitem[\protect\citeauthoryear{{Carvajal}, {Favre}, {Kleiner}, {Ceccarelli},
  {Bergin}  \& {Fedele}}{{Carvajal} et~al.}{2019}]{Carvajal2019}
{Carvajal} M.,  {Favre} C.,  {Kleiner} I.,  {Ceccarelli} C.,  {Bergin} E.~A.,
  {Fedele} D.,  2019, \mn@doi [\aap] {10.1051/0004-6361/201935469}, \href
  {https://ui.adsabs.harvard.edu/abs/2019A&A...627A..65C} {627, A65}

\bibitem[\protect\citeauthoryear{{Churchwell}, {Wood}, {Myers}  \&
  {Myers}}{{Churchwell} et~al.}{1986}]{Churchwell1986}
{Churchwell} E.,  {Wood} D.,  {Myers} P.~C.,   {Myers} R.~V.,  1986, \mn@doi
  [\apj] {10.1086/164256}, \href
  {https://ui.adsabs.harvard.edu/abs/1986ApJ...305..405C} {305, 405}

\bibitem[\protect\citeauthoryear{{Crovisier}}{{Crovisier}}{1998}]{Crovisier98}
{Crovisier} J.,  1998, \mn@doi [Faraday Discussions] {10.1039/a800079d}, \href
  {https://ui.adsabs.harvard.edu/abs/1998FaDi..109..437C} {109, 437}

\bibitem[\protect\citeauthoryear{{Dalgarno}}{{Dalgarno}}{2006}]{Dalgarno2006}
{Dalgarno} A.,  2006, \mn@doi [Proceedings of the National Academy of Science]
  {10.1073/pnas.0602117103}, \href
  {https://ui.adsabs.harvard.edu/abs/2006PNAS..10312269D} {103, 12269}

\bibitem[\protect\citeauthoryear{{Dobrijevic}, {H{\'e}brard}, {Loison}  \&
  {Hickson}}{{Dobrijevic} et~al.}{2014}]{Dob14}
{Dobrijevic} M.,  {H{\'e}brard} E.,  {Loison} J.~C.,   {Hickson} K.~M.,  2014,
  \mn@doi [\icarus] {10.1016/j.icarus.2013.10.015}, \href
  {https://ui.adsabs.harvard.edu/abs/2014Icar..228..324D} {228, 324}

\bibitem[\protect\citeauthoryear{{Dumke} \& {Mac-Auliffe}}{{Dumke} \&
  {Mac-Auliffe}}{2010}]{dum2010}
{Dumke} M.,  {Mac-Auliffe} F.,  2010, in \procspie. p. 77371J,
  \mn@doi{10.1117/12.858020}

\bibitem[\protect\citeauthoryear{{Duronea} et~al.,}{{Duronea}
  et~al.}{2019}]{Duronea19}
{Duronea} N.~U.,  et~al., 2019, \mn@doi [\mnras] {10.1093/mnras/stz2087}, \href
  {https://ui.adsabs.harvard.edu/abs/2019MNRAS.489.1519D} {489, 1519}

\bibitem[\protect\citeauthoryear{East, Johnson  \& Allen}{East
  et~al.}{1993}]{East1993}
East A.~L.~L.,  Johnson C.~S.,   Allen W.~D.,  1993, J.\ Chem.\ Phys., 98, 1299

\bibitem[\protect\citeauthoryear{{Endres}, {Schlemmer}, {Schilke}, {Stutzki}
  \& {M{\"u}ller}}{{Endres} et~al.}{2016}]{end2016}
{Endres} C.~P.,  {Schlemmer} S.,  {Schilke} P.,  {Stutzki} J.,   {M{\"u}ller}
  H.~S.~P.,  2016, \mn@doi [Journal of Molecular Spectroscopy]
  {10.1016/j.jms.2016.03.005}, \href
  {https://ui.adsabs.harvard.edu/abs/2016JMoSp.327...95E} {327, 95}

\bibitem[\protect\citeauthoryear{Favre et~al.,}{Favre et~al.}{2014}]{Favre2014}
Favre C.,  et~al., 2014, \mn@doi [\apjs] {10.1088/0067-0049/215/2/25}, 215, 25

\bibitem[\protect\citeauthoryear{{Fedoseev}, {Ioppolo}, {Zhao}, {Lamberts}  \&
  {Linnartz}}{{Fedoseev} et~al.}{2015}]{Fedoseev15}
{Fedoseev} G.,  {Ioppolo} S.,  {Zhao} D.,  {Lamberts} T.,   {Linnartz} H.,
  2015, \mn@doi [\mnras] {10.1093/mnras/stu2028}, \href
  {https://ui.adsabs.harvard.edu/abs/2015MNRAS.446..439F} {446, 439}

\bibitem[\protect\citeauthoryear{{Ghesqui{\`e}re}, {Mineva}, {Talbi},
  {Theul{\'e}}, {Noble}  \& {Chiavassa}}{{Ghesqui{\`e}re} et~al.}{2015}]{ghe15}
{Ghesqui{\`e}re} P.,  {Mineva} T.,  {Talbi} D.,  {Theul{\'e}} P.,  {Noble}
  J.~A.,   {Chiavassa} T.,  2015, \mn@doi [Physical Chemistry Chemical Physics
  (Incorporating Faraday Transactions)] {10.1039/C5CP00558B}, \href
  {http://adsabs.harvard.edu/abs/2015PCCP...1711455G} {17, 11455}

\bibitem[\protect\citeauthoryear{{Goldsmith} \& {Langer}}{{Goldsmith} \&
  {Langer}}{1999}]{gol1999}
{Goldsmith} P.~F.,  {Langer} W.~D.,  1999, \mn@doi [\apj] {10.1086/307195},
  \href {https://ui.adsabs.harvard.edu/abs/1999ApJ...517..209G} {517, 209}

\bibitem[\protect\citeauthoryear{{Gorai}, {Bhat}, {Sil}, {Mondal}, {Ghosh},
  {Chakrabarti}  \& {Das}}{{Gorai} et~al.}{2020}]{Gorai2020}
{Gorai} P.,  {Bhat} B.,  {Sil} M.,  {Mondal} S.~K.,  {Ghosh} R.,  {Chakrabarti}
  S.~K.,   {Das} A.,  2020, \mn@doi [\apj] {10.3847/1538-4357/ab8871}, \href
  {https://ui.adsabs.harvard.edu/abs/2020ApJ...895...86G} {895, 86}

\bibitem[\protect\citeauthoryear{{Graedel}, {Langer}  \& {Frerking}}{{Graedel}
  et~al.}{1982}]{Graedel82}
{Graedel} T.~E.,  {Langer} W.~D.,   {Frerking} M.~A.,  1982, \mn@doi [\apjs]
  {10.1086/190780}, \href
  {https://ui.adsabs.harvard.edu/abs/1982ApJS...48..321G} {48, 321}

\bibitem[\protect\citeauthoryear{{Greaves} \& {White}}{{Greaves} \&
  {White}}{1991}]{Gre91}
{Greaves} J.~S.,  {White} G.~J.,  1991, \aaps, \href
  {https://ui.adsabs.harvard.edu/abs/1991A&AS...91..237G} {91, 237}

\bibitem[\protect\citeauthoryear{{G{\"u}sten}, {Nyman}, {Schilke}, {Menten},
  {Cesarsky}  \& {Booth}}{{G{\"u}sten} et~al.}{2006}]{gu06}
{G{\"u}sten} R.,  {Nyman} L.~{\AA}.,  {Schilke} P.,  {Menten} K.,  {Cesarsky}
  C.,   {Booth} R.,  2006, \mn@doi [\aap] {10.1051/0004-6361:20065420}, \href
  {http://adsabs.harvard.edu/abs/2006A%26A...454L..13G} {454, L13}

\bibitem[\protect\citeauthoryear{{Hasegawa} \& {Herbst}}{{Hasegawa} \&
  {Herbst}}{1993}]{has93}
{Hasegawa} T.~I.,  {Herbst} E.,  1993, \mn@doi [\mnras]
  {10.1093/mnras/263.3.589}, \href
  {http://adsabs.harvard.edu/abs/1993MNRAS.263..589H} {263, 589}

\bibitem[\protect\citeauthoryear{{He} et~al.,}{{He} et~al.}{2021}]{He2021}
{He} Y.-X.,  et~al., 2021, \mn@doi [\apjs] {10.3847/1538-4365/abd0fb}, \href
  {https://ui.adsabs.harvard.edu/abs/2021ApJS..253....2H} {253, 2}

\bibitem[\protect\citeauthoryear{{Herbst} \& {van Dishoeck}}{{Herbst} \& {van
  Dishoeck}}{2009}]{Herbst2009}
{Herbst} E.,  {van Dishoeck} E.~F.,  2009, \mn@doi [\araa]
  {10.1146/annurev-astro-082708-101654}, \href
  {https://ui.adsabs.harvard.edu/abs/2009ARA&A..47..427H} {47, 427}

\bibitem[\protect\citeauthoryear{{Hern{\'a}ndez-G{\'o}mez}, {Sahnoun}, {Caux},
  {Wiesenfeld}, {Loinard}, {Bottinelli}, {Hammami}  \&
  {Menten}}{{Hern{\'a}ndez-G{\'o}mez} et~al.}{2019}]{Hernandez19}
{Hern{\'a}ndez-G{\'o}mez} A.,  {Sahnoun} E.,  {Caux} E.,  {Wiesenfeld} L.,
  {Loinard} L.,  {Bottinelli} S.,  {Hammami} K.,   {Menten} K.~M.,  2019,
  \mn@doi [\mnras] {10.1093/mnras/sty2971}, \href
  {https://ui.adsabs.harvard.edu/abs/2019MNRAS.483.2014H} {483, 2014}

\bibitem[\protect\citeauthoryear{{Herv{\'\i}as-Caimapo}
  et~al.,}{{Herv{\'\i}as-Caimapo} et~al.}{2019}]{Hervias19}
{Herv{\'\i}as-Caimapo} C.,  et~al., 2019, \mn@doi [\apj]
  {10.3847/1538-4357/aaf9ac}, \href
  {https://ui.adsabs.harvard.edu/abs/2019ApJ...872..200H} {872, 200}

\bibitem[\protect\citeauthoryear{Herzberg}{Herzberg}{1991}]{Herzberg}
Herzberg G.,  1991, Spectra and Molecular Structure: II. Infrared and Raman
  Spectra of Polyatomic Molecules.
Krieger Pub. Co., Malabar, Florida

\bibitem[\protect\citeauthoryear{{Hincelin}, {Wakelam}, {Hersant},
  {Guilloteau}, {Loison}, {Honvault}  \& {Troe}}{{Hincelin}
  et~al.}{2011}]{Hincelin11}
{Hincelin} U.,  {Wakelam} V.,  {Hersant} F.,  {Guilloteau} S.,  {Loison} J.~C.,
   {Honvault} P.,   {Troe} J.,  2011, \mn@doi [\aap]
  {10.1051/0004-6361/201016328}, \href
  {https://ui.adsabs.harvard.edu/abs/2011A&A...530A..61H} {530, A61}

\bibitem[\protect\citeauthoryear{{Hocking}, {Gerry}  \&
  {Winnewisser}}{{Hocking} et~al.}{1972}]{Hocking1972}
{Hocking} W.~H.,  {Gerry} M.~C.~L.,   {Winnewisser} G.,  1972, \mn@doi [\apjl]
  {10.1086/180956}, \href
  {https://ui.adsabs.harvard.edu/abs/1972ApJ...174L..93H} {174, L93}

\bibitem[\protect\citeauthoryear{{Hocking}, {Gerry}  \&
  {Winnewisser}}{{Hocking} et~al.}{1974}]{Hocking1974}
{Hocking} W.~H.,  {Gerry} M.~C.~L.,   {Winnewisser} G.,  1974, \mn@doi [\apjl]
  {10.1086/181403}, \href
  {https://ui.adsabs.harvard.edu/abs/1974ApJ...187L..89H} {187, L89}

\bibitem[\protect\citeauthoryear{Hocking, Gerry  \& Winnewisser}{Hocking
  et~al.}{1975}]{Hocking1975}
Hocking W.~H.,  Gerry M. C.~L.,   Winnewisser G.,  1975, \mn@doi [Canadian
  Journal of Physics] {10.1139/p75-239}, 53, 1869

\bibitem[\protect\citeauthoryear{{Jenkins}}{{Jenkins}}{2009}]{Jenkins09}
{Jenkins} E.~B.,  2009, \mn@doi [\apj] {10.1088/0004-637X/700/2/1299}, \href
  {https://ui.adsabs.harvard.edu/abs/2009ApJ...700.1299J} {700, 1299}

\bibitem[\protect\citeauthoryear{Jensen}{Jensen}{1983}]{Jensen1983}
Jensen P.,  1983, J. Mol. Spectrosc., 101, 422

\bibitem[\protect\citeauthoryear{{Jewell}, {Hollis}, {Lovas}  \&
  {Snyder}}{{Jewell} et~al.}{1989}]{Jew89}
{Jewell} P.~R.,  {Hollis} J.~M.,  {Lovas} F.~J.,   {Snyder} L.~E.,  1989,
  \mn@doi [\apjs] {10.1086/191359}, \href
  {https://ui.adsabs.harvard.edu/abs/1989ApJS...70..833J} {70, 833}

\bibitem[\protect\citeauthoryear{{Jones} \& {Badger}}{{Jones} \&
  {Badger}}{1950}]{Jones1950b}
{Jones} L.~H.,  {Badger} R.~M.,  1950, \mn@doi [\jcp] {10.1063/1.1747523},
  \href {https://ui.adsabs.harvard.edu/abs/1950JChPh..18.1511J} {18, 1511}

\bibitem[\protect\citeauthoryear{{Jones}, {Shoolery}, {Shulman}  \&
  {Yost}}{{Jones} et~al.}{1950}]{Jones1950a}
{Jones} L.~H.,  {Shoolery} J.~N.,  {Shulman} R.~G.,   {Yost} D.~M.,  1950,
  \mn@doi [\jcp] {10.1063/1.1747827}, \href
  {https://ui.adsabs.harvard.edu/abs/1950JChPh..18..990J} {18, 990}

\bibitem[\protect\citeauthoryear{Khalouf-Rivera, Pérez-Bernal  \&
  Carvajal}{Khalouf-Rivera et~al.}{2020}]{KhaloufRivera2020}
Khalouf-Rivera J.,  Pérez-Bernal F.,   Carvajal M.,  2020, \mn@doi [Journal of
  Quantitative Spectroscopy and Radiative Transfer]
  {https://doi.org/10.1016/j.jqsrt.2020.107436}, p. 107436

\bibitem[\protect\citeauthoryear{Lapinov, Golubiatnikov, Markov  \&
  Guarnieri}{Lapinov et~al.}{2007}]{Lapinov2007}
Lapinov A.~V.,  Golubiatnikov G.~Y.,  Markov V.~N.,   Guarnieri A.,  2007,
  Astron.\ Lett., 33, 121

\bibitem[\protect\citeauthoryear{{Li}, {Wang}, {Gu}  \& {Zheng}}{{Li}
  et~al.}{2013}]{Li2013}
{Li} J.,  {Wang} J.~Z.,  {Gu} Q.~S.,   {Zheng} X.~W.,  2013, \mn@doi [\aap]
  {10.1051/0004-6361/201220943}, \href
  {https://ui.adsabs.harvard.edu/abs/2013A&A...555A..18L} {555, A18}

\bibitem[\protect\citeauthoryear{{Ligterink}, {Terwisscha van Scheltinga},
  {Taquet}, {J{\o}rgensen}, {Cazaux}, {van Dishoeck}  \&
  {Linnartz}}{{Ligterink} et~al.}{2018}]{Ligterink18}
{Ligterink} N.~F.~W.,  {Terwisscha van Scheltinga} J.,  {Taquet} V.,
  {J{\o}rgensen} J.~K.,  {Cazaux} S.,  {van Dishoeck} E.~F.,   {Linnartz} H.,
  2018, \mn@doi [\mnras] {10.1093/mnras/sty2066}, \href
  {https://ui.adsabs.harvard.edu/abs/2018MNRAS.480.3628L} {480, 3628}

\bibitem[\protect\citeauthoryear{{Lis} et~al.,}{{Lis} et~al.}{1997}]{Lis1997}
{Lis} D.~C.,  et~al., 1997, \mn@doi [\icarus] {10.1006/icar.1997.5833}, \href
  {https://ui.adsabs.harvard.edu/abs/1997Icar..130..355L} {130, 355}

\bibitem[\protect\citeauthoryear{{L{\'o}pez-Sepulcre}
  et~al.,}{{L{\'o}pez-Sepulcre} et~al.}{2015}]{Lopez2015}
{L{\'o}pez-Sepulcre} A.,  et~al., 2015, \mn@doi [\mnras]
  {10.1093/mnras/stv377}, \href
  {https://ui.adsabs.harvard.edu/abs/2015MNRAS.449.2438L} {449, 2438}

\bibitem[\protect\citeauthoryear{{L{\'o}pez-Sepulcre}, {Balucani},
  {Ceccarelli}, {Codella}, {Dulieu}  \& {Theul{\'e}}}{{L{\'o}pez-Sepulcre}
  et~al.}{2019}]{Lopez2019}
{L{\'o}pez-Sepulcre} A.,  {Balucani} N.,  {Ceccarelli} C.,  {Codella} C.,
  {Dulieu} F.,   {Theul{\'e}} P.,  2019, \mn@doi [ACS Earth and Space
  Chemistry] {10.1021/acsearthspacechem.9b00154}, \href
  {https://ui.adsabs.harvard.edu/abs/2019ECS.....3.2122L} {3, 2122}

\bibitem[\protect\citeauthoryear{{Loren} \& {Mundy}}{{Loren} \&
  {Mundy}}{1984}]{Lor84}
{Loren} R.~B.,  {Mundy} L.~G.,  1984, \mn@doi [\apj] {10.1086/162591}, \href
  {https://ui.adsabs.harvard.edu/abs/1984ApJ...286..232L} {286, 232}

\bibitem[\protect\citeauthoryear{{Lovas}}{{Lovas}}{2004}]{Lovas2004}
{Lovas} F.~J.,  2004, \mn@doi [Journal of Physical and Chemical Reference Data]
  {10.1063/1.1633275}, \href
  {https://ui.adsabs.harvard.edu/abs/2004JPCRD..33..177L} {33, 177}

\bibitem[\protect\citeauthoryear{{MacDonald}, {Gibb}, {Habing}  \&
  {Millar}}{{MacDonald} et~al.}{1996}]{Mac96}
{MacDonald} G.~H.,  {Gibb} A.~G.,  {Habing} R.~J.,   {Millar} T.~J.,  1996,
  \aaps, \href {https://ui.adsabs.harvard.edu/abs/1996A&AS..119..333M} {119,
  333}

\bibitem[\protect\citeauthoryear{{Marcelino}, {Cernicharo}, {Tercero}  \&
  {Roueff}}{{Marcelino} et~al.}{2009}]{Marcelino2009}
{Marcelino} N.,  {Cernicharo} J.,  {Tercero} B.,   {Roueff} E.,  2009, \mn@doi
  [\apjl] {10.1088/0004-637X/690/1/L27}, \href
  {https://ui.adsabs.harvard.edu/abs/2009ApJ...690L..27M} {690, L27}

\bibitem[\protect\citeauthoryear{{Mart{\'\i}n-Dom{\'e}nech}, {Rivilla},
  {Jim{\'e}nez-Serra}, {Qu{\'e}nard}, {Testi}  \&
  {Mart{\'\i}n-Pintado}}{{Mart{\'\i}n-Dom{\'e}nech} et~al.}{2017}]{Martin17}
{Mart{\'\i}n-Dom{\'e}nech} R.,  {Rivilla} V.~M.,  {Jim{\'e}nez-Serra} I.,
  {Qu{\'e}nard} D.,  {Testi} L.,   {Mart{\'\i}n-Pintado} J.,  2017, \mn@doi
  [\mnras] {10.1093/mnras/stx915}, \href
  {https://ui.adsabs.harvard.edu/abs/2017MNRAS.469.2230M} {469, 2230}

\bibitem[\protect\citeauthoryear{{Mart{\'\i}n}, {Mauersberger},
  {Mart{\'\i}n-Pintado}, {Henkel}  \& {Garc{\'\i}a-Burillo}}{{Mart{\'\i}n}
  et~al.}{2006}]{Martin06}
{Mart{\'\i}n} S.,  {Mauersberger} R.,  {Mart{\'\i}n-Pintado} J.,  {Henkel} C.,
   {Garc{\'\i}a-Burillo} S.,  2006, \mn@doi [\apjs] {10.1086/503297}, \href
  {https://ui.adsabs.harvard.edu/abs/2006ApJS..164..450M} {164, 450}

\bibitem[\protect\citeauthoryear{{Mart{\'\i}n}, {Requena-Torres},
  {Mart{\'\i}n-Pintado}  \& {Mauersberger}}{{Mart{\'\i}n}
  et~al.}{2008}]{Martin2008}
{Mart{\'\i}n} S.,  {Requena-Torres} M.~A.,  {Mart{\'\i}n-Pintado} J.,
  {Mauersberger} R.,  2008, \mn@doi [\apj] {10.1086/533409}, \href
  {https://ui.adsabs.harvard.edu/abs/2008ApJ...678..245M} {678, 245}

\bibitem[\protect\citeauthoryear{{Mart{\'\i}n}, {Mart{\'\i}n-Pintado}  \&
  {Mauersberger}}{{Mart{\'\i}n} et~al.}{2009}]{Martin09}
{Mart{\'\i}n} S.,  {Mart{\'\i}n-Pintado} J.,   {Mauersberger} R.,  2009,
  \mn@doi [\apj] {10.1088/0004-637X/694/1/610}, \href
  {https://ui.adsabs.harvard.edu/abs/2009ApJ...694..610M} {694, 610}

\bibitem[\protect\citeauthoryear{McLean, Loew  \& Berkowitz}{McLean
  et~al.}{1977}]{McLean1977}
McLean A.,  Loew G.,   Berkowitz D.,  1977, J. Mol. Spectrosc., 64, 184

\bibitem[\protect\citeauthoryear{{Meier} \& {Turner}}{{Meier} \&
  {Turner}}{2005}]{Meier05}
{Meier} D.~S.,  {Turner} J.~L.,  2005, \mn@doi [\apj] {10.1086/426499}, \href
  {https://ui.adsabs.harvard.edu/abs/2005ApJ...618..259M} {618, 259}

\bibitem[\protect\citeauthoryear{{Mendoza}, {Lefloch}, {L{\'o}pez-Sepulcre},
  {Ceccarelli}, {Codella}, {Boechat-Roberty}  \& {Bachiller}}{{Mendoza}
  et~al.}{2014}]{Mendoza14}
{Mendoza} E.,  {Lefloch} B.,  {L{\'o}pez-Sepulcre} A.,  {Ceccarelli} C.,
  {Codella} C.,  {Boechat-Roberty} H.~M.,   {Bachiller} R.,  2014, \mn@doi
  [\mnras] {10.1093/mnras/stu1718}, \href
  {https://ui.adsabs.harvard.edu/abs/2014MNRAS.445..151M} {445, 151}

\bibitem[\protect\citeauthoryear{{Mendoza} et~al.,}{{Mendoza}
  et~al.}{2018}]{Mendoza18}
{Mendoza} E.,  et~al., 2018, \mn@doi [\apj] {10.3847/1538-4357/aaa1ec}, \href
  {https://ui.adsabs.harvard.edu/abs/2018ApJ...853..152M} {853, 152}

\bibitem[\protect\citeauthoryear{{Merello}, {Bronfman}, {Garay}, {Nyman},
  {Evans}  \& {Walmsley}}{{Merello} et~al.}{2013a}]{Merello13b}
{Merello} M.,  {Bronfman} L.,  {Garay} G.,  {Nyman} L.-{\r{A}}.,  {Evans}
  Neal~J. I.,   {Walmsley} C.~M.,  2013a, \mn@doi [\apj]
  {10.1088/0004-637X/774/1/38}, \href
  {https://ui.adsabs.harvard.edu/abs/2013ApJ...774...38M} {774, 38}

\bibitem[\protect\citeauthoryear{{Merello}, {Bronfman}, {Garay}, {Lo}, {Evans},
  {Nyman}, {Cort{\'e}s}  \& {Cunningham}}{{Merello} et~al.}{2013b}]{Merello13a}
{Merello} M.,  {Bronfman} L.,  {Garay} G.,  {Lo} N.,  {Evans} Neal~J. I.,
  {Nyman} L.-{\r{A}}.,  {Cort{\'e}s} J.~R.,   {Cunningham} M.~R.,  2013b,
  \mn@doi [\apjl] {10.1088/2041-8205/774/1/L7}, \href
  {https://ui.adsabs.harvard.edu/abs/2013ApJ...774L...7M} {774, L7}

\bibitem[\protect\citeauthoryear{Mladenovic, Lewerenz, McCarthy  \&
  Thaddeus}{Mladenovic et~al.}{2009}]{Mladenovic2009}
Mladenovic M.,  Lewerenz M.,  McCarthy M.,   Thaddeus P.,  2009, J. Chem.
  Phys., 131, 174308

\bibitem[\protect\citeauthoryear{Mladenović \& Lewerenz}{Mladenović \&
  Lewerenz}{2008}]{Mladenovic2008}
Mladenović M.,  Lewerenz M.,  2008, \mn@doi [Chemical Physics]
  {https://doi.org/10.1016/j.chemphys.2007.06.033}, 343, 129

\bibitem[\protect\citeauthoryear{{Neufeld}, {Wolfire}  \& {Schilke}}{{Neufeld}
  et~al.}{2005}]{Neufeld05}
{Neufeld} D.~A.,  {Wolfire} M.~G.,   {Schilke} P.,  2005, \mn@doi [\apj]
  {10.1086/430663}, \href
  {https://ui.adsabs.harvard.edu/abs/2005ApJ...628..260N} {628, 260}

\bibitem[\protect\citeauthoryear{{Nguyen-Q-Rieu}, {Henkel}, {Jackson}  \&
  {Mauersberger}}{{Nguyen-Q-Rieu} et~al.}{1991}]{Nguyen1991}
{Nguyen-Q-Rieu} {Henkel} C.,  {Jackson} J.~M.,   {Mauersberger} R.,  1991,
  \aap, \href {https://ui.adsabs.harvard.edu/abs/1991A&A...241L..33N} {241,
  L33}

\bibitem[\protect\citeauthoryear{{Niedenhoff}, {Yamada}, {Belov}  \&
  {Winnewisser}}{{Niedenhoff} et~al.}{1995}]{Niedenhoff1995}
{Niedenhoff} M.,  {Yamada} K.~M.~T.,  {Belov} S.~P.,   {Winnewisser} G.,  1995,
  \mn@doi [Journal of Molecular Spectroscopy] {10.1006/jmsp.1995.1277}, \href
  {https://ui.adsabs.harvard.edu/abs/1995JMoSp.174..151N} {174, 151}

\bibitem[\protect\citeauthoryear{{Nummelin}, {Bergman}, {Hjalmarson},
  {Friberg}, {Irvine}, {Millar}, {Ohishi}  \& {Saito}}{{Nummelin}
  et~al.}{1998}]{Nummelin1998}
{Nummelin} A.,  {Bergman} P.,  {Hjalmarson} {\r{A}}.,  {Friberg} P.,  {Irvine}
  W.~M.,  {Millar} T.~J.,  {Ohishi} M.,   {Saito} S.,  1998, \mn@doi [\apjs]
  {10.1086/313126}, \href
  {https://ui.adsabs.harvard.edu/abs/1998ApJS..117..427N} {117, 427}

\bibitem[\protect\citeauthoryear{Pickett}{Pickett}{1991}]{Pickett1991}
Pickett H.~M.,  1991, J. Mol. Spectr., 148, 371

\bibitem[\protect\citeauthoryear{{Pickett}, {Poynter}, {Cohen}, {Delitsky},
  {Pearson}  \& {M{\"u}ller}}{{Pickett} et~al.}{1998}]{pic1998}
{Pickett} H.~M.,  {Poynter} R.~L.,  {Cohen} E.~A.,  {Delitsky} M.~L.,
  {Pearson} J.~C.,   {M{\"u}ller} H.~S.~P.,  1998, \mn@doi [\jqsrt]
  {10.1016/S0022-4073(98)00091-0}, \href
  {https://ui.adsabs.harvard.edu/abs/1998JQSRT..60..883P} {60, 883}

\bibitem[\protect\citeauthoryear{{Prasad} \& {Tarafdar}}{{Prasad} \&
  {Tarafdar}}{1983}]{Prasad83}
{Prasad} S.~S.,  {Tarafdar} S.~P.,  1983, \mn@doi [\apj] {10.1086/160896},
  \href {https://ui.adsabs.harvard.edu/abs/1983ApJ...267..603P} {267, 603}

\bibitem[\protect\citeauthoryear{{Reboussin}, {Wakelam}, {Guilloteau}  \&
  {Hersant}}{{Reboussin} et~al.}{2014}]{reb14}
{Reboussin} L.,  {Wakelam} V.,  {Guilloteau} S.,   {Hersant} F.,  2014, \mn@doi
  [\mnras] {10.1093/mnras/stu462}, \href
  {http://adsabs.harvard.edu/abs/2014MNRAS.440.3557R} {440, 3557}

\bibitem[\protect\citeauthoryear{{Ruaud}, {Loison}, {Hickson}, {Gratier},
  {Hersant}  \& {Wakelam}}{{Ruaud} et~al.}{2015}]{rua15}
{Ruaud} M.,  {Loison} J.~C.,  {Hickson} K.~M.,  {Gratier} P.,  {Hersant} F.,
  {Wakelam} V.,  2015, \mn@doi [\mnras] {10.1093/mnras/stu2709}, \href
  {http://adsabs.harvard.edu/abs/2015MNRAS.447.4004R} {447, 4004}

\bibitem[\protect\citeauthoryear{{Ruaud}, {Wakelam}  \& {Hersant}}{{Ruaud}
  et~al.}{2016}]{nautilus}
{Ruaud} M.,  {Wakelam} V.,   {Hersant} F.,  2016, \mn@doi [MNRAS]
  {10.1093/mnras/stw887}, \href
  {http://adsabs.harvard.edu/abs/2016MNRAS.459.3756R} {459, 3756}

\bibitem[\protect\citeauthoryear{Schulze, Koja, Winnewisser  \&
  Winnewisser}{Schulze et~al.}{2000}]{Schulze2000}
Schulze G.,  Koja O.,  Winnewisser B.,   Winnewisser M.,  2000, J. Mol.
  Structure, 517/518, 307

\bibitem[\protect\citeauthoryear{{Semenov} et~al.,}{{Semenov}
  et~al.}{2010}]{sem10}
{Semenov} D.,  et~al., 2010, \mn@doi [\aap] {10.1051/0004-6361/201015149},
  \href {http://adsabs.harvard.edu/abs/2010A%26A...522A..42S} {522, A42}

\bibitem[\protect\citeauthoryear{Shimanouchi}{Shimanouchi}{2018}]{NIST}
Shimanouchi T.,  2018, Molecular Vibrational Frequencies in NIST Chemistry
  WebBook. NIST Standard Reference Database Number 69.
P.J. Linstrom and W.G. Mallard, National Institute of Standards and Technology,
  Gaithersburg MD, 20899, USA, \mn@doi{https://doi.org/10.18434/T4D303}

\bibitem[\protect\citeauthoryear{{Snyder} \& {Buhl}}{{Snyder} \&
  {Buhl}}{1972}]{snyder72}
{Snyder} L.~E.,  {Buhl} D.,  1972, \apj, 177, 619

\bibitem[\protect\citeauthoryear{{Snyder} et~al.,}{{Snyder}
  et~al.}{2005}]{Snyder2005}
{Snyder} L.~E.,  et~al., 2005, \mn@doi [\apj] {10.1086/426677}, \href
  {https://ui.adsabs.harvard.edu/abs/2005ApJ...619..914S} {619, 914}

\bibitem[\protect\citeauthoryear{Su, Kong, Chen, Huang  \& Liu}{Su
  et~al.}{2000}]{Su2000}
Su H.,  Kong F.,  Chen B.,  Huang M.,   Liu Y.,  2000, J. Chem. Phys., 113,
  1885

\bibitem[\protect\citeauthoryear{{Sutton}, {Blake}, {Masson}  \&
  {Phillips}}{{Sutton} et~al.}{1985}]{Sut85}
{Sutton} E.~C.,  {Blake} G.~A.,  {Masson} C.~R.,   {Phillips} T.~G.,  1985,
  \mn@doi [\apjs] {10.1086/191045}, \href
  {https://ui.adsabs.harvard.edu/abs/1985ApJS...58..341S} {58, 341}

\bibitem[\protect\citeauthoryear{{Sutton}, {Jaminet}, {Danchi}  \&
  {Blake}}{{Sutton} et~al.}{1991}]{Sut91}
{Sutton} E.~C.,  {Jaminet} P.~A.,  {Danchi} W.~C.,   {Blake} G.~A.,  1991,
  \mn@doi [\apjs] {10.1086/191603}, \href
  {https://ui.adsabs.harvard.edu/abs/1991ApJS...77..255S} {77, 255}

\bibitem[\protect\citeauthoryear{Takashi, Tanaka  \& Tanaka}{Takashi
  et~al.}{1989}]{Takashi1989}
Takashi R.,  Tanaka K.,   Tanaka T.,  1989, \mn@doi [Journal of Molecular
  Spectroscopy] {https://doi.org/10.1016/0022-2852(89)90012-X}, 138, 450

\bibitem[\protect\citeauthoryear{{Turner}}{{Turner}}{1991}]{Turner91}
{Turner} B.~E.,  1991, \mn@doi [\apjs] {10.1086/191577}, \href
  {https://ui.adsabs.harvard.edu/abs/1991ApJS...76..617T} {76, 617}

\bibitem[\protect\citeauthoryear{{Van der Tak} \& {van Dishoeck}}{{Van der Tak}
  \& {van Dishoeck}}{2000}]{vandertak2000}
{Van der Tak} F.~F.~S.,  {van Dishoeck} E.~F.,  2000, \aap, \href
  {https://ui.adsabs.harvard.edu/abs/2000A&A...358L..79V} {358, L79}

\bibitem[\protect\citeauthoryear{{Vassilev} et~al.,}{{Vassilev}
  et~al.}{2008}]{vas08}
{Vassilev} V.,  et~al., 2008, \mn@doi [\aap] {10.1051/0004-6361:200810459},
  \href {http://adsabs.harvard.edu/abs/2008A%26A...490.1157V} {490, 1157}

\bibitem[\protect\citeauthoryear{{Velilla Prieto} et~al.,}{{Velilla Prieto}
  et~al.}{2015}]{Prieto15}
{Velilla Prieto} L.,  et~al., 2015, \mn@doi [\aap]
  {10.1051/0004-6361/201424768}, \href
  {https://ui.adsabs.harvard.edu/abs/2015A&A...575A..84V} {575, A84}

\bibitem[\protect\citeauthoryear{{Vidal} \& {Wakelam}}{{Vidal} \&
  {Wakelam}}{2018}]{vid18}
{Vidal} T.~H.~G.,  {Wakelam} V.,  2018, \mn@doi [MNRAS]
  {10.1093/mnras/stx3113}, \href
  {http://adsabs.harvard.edu/abs/2018MNRAS.474.5575V} {474, 5575}

\bibitem[\protect\citeauthoryear{{Wakelam} \& {Herbst}}{{Wakelam} \&
  {Herbst}}{2008}]{wak08}
{Wakelam} V.,  {Herbst} E.,  2008, \mn@doi [\apj] {10.1086/587734}, \href
  {https://ui.adsabs.harvard.edu/abs/2008ApJ...680..371W} {680, 371}

\bibitem[\protect\citeauthoryear{{Wakelam}, {Herbst}  \& {Selsis}}{{Wakelam}
  et~al.}{2006}]{wak06}
{Wakelam} V.,  {Herbst} E.,   {Selsis} F.,  2006, \mn@doi [\aap]
  {10.1051/0004-6361:20054682}, \href
  {https://ui.adsabs.harvard.edu/abs/2006A&A...451..551W} {451, 551}

\bibitem[\protect\citeauthoryear{{Wakelam} et~al.,}{{Wakelam}
  et~al.}{2015}]{Wak15}
{Wakelam} V.,  et~al., 2015, \mn@doi [ApJS] {10.1088/0067-0049/217/2/20}, \href
  {https://ui.adsabs.harvard.edu/abs/2015ApJS..217...20W} {217, 20}

\bibitem[\protect\citeauthoryear{{Wells} \& {Raston}}{{Wells} \&
  {Raston}}{2020}]{Wells2020}
{Wells} T.,  {Raston} P.,  2020, J. Mol. Spectroscopy, 370, 111292

\bibitem[\protect\citeauthoryear{Winnewisser \& Winnewisser}{Winnewisser \&
  Winnewisser}{1971}]{Winnewisser1971}
Winnewisser M.,  Winnewisser B.,  1971, Z. Naturforsch., 26, 128

\bibitem[\protect\citeauthoryear{{Wyrowski}, {Schilke}  \&
  {Walmsley}}{{Wyrowski} et~al.}{1999}]{Wyrowski1999}
{Wyrowski} F.,  {Schilke} P.,   {Walmsley} C.~M.,  1999, \aap, \href
  {https://ui.adsabs.harvard.edu/abs/1999A&A...341..882W} {341, 882}

\bibitem[\protect\citeauthoryear{{Zinchenko}, {Henkel}  \& {Mao}}{{Zinchenko}
  et~al.}{2000}]{Zinchenko2000}
{Zinchenko} I.,  {Henkel} C.,   {Mao} R.~Q.,  2000, \aap, \href
  {https://ui.adsabs.harvard.edu/abs/2000A&A...361.1079Z} {361, 1079}

\makeatother
\end{thebibliography}


\begin{appendix}

\section{Calculation of the partition functions of CHNO isomers}
\label{appendix-Qrv}

Here we are giving some details for the calculation of the rovibrational partition function for the isomers of CHNO isomers (isocyanic acid, cyanic acid, fulminic acid and isofulminic acid).  

\subsection{Isocyanic acid (HNCO)}
\label{Q-HNCO}

Isocyanic acid is a near-prolate asymmetric top of which  rotational-vibrational partition function has been previously calculated \citep{Carvajal2019}. The main results of this  isomer have been included here to compare them with the internal partition functions of the other three isomers. Thus, here we only outline the details for its calculation.

The internal partition function has been computed from Eq.~\ref{eq:rtvQapprox}. The rotational partition function has been computed as a direct sum taking into consideration the rotational energies predicted up to $J$~=~135 and K$_a$~=~30 from the Hamiltonian provided by \cite{Lapinov2007}. The harmonic vibrational partition function has been calculated with the experimental vibrational fundamental frequencies recorded by~\cite{East1993}.

\subsection{Cyanic acid (HOCN)}
\label{Q-HOCN}

Cyanic acid is a bent molecule with a near-prolate asymmetric top structure of which rotational spectrum has been measured in the centimeter- and millimeter-wave windows~\citep{Brunken2009}. This molecular species has no available experimental vibrational energies in the gas phase except the $\nu_2$ band~\citep{Su2000,NIST} but the complete set of fundamental frequencies have been obtained theoretically~\citep{Mladenovic2008}.

The values of the rotational partition function calculated as a direct sum and as a classical approximation are given in Table~\ref{tab-Qrot-HOCN}. The direct sum is obtained using the predicted rotational energy levels up to $J$~=~72 and K$_a$=8~\citep{Brunken2009} provided in the CDMS ~\citep{end2016}. The classical approximation is computed with the rotational constants from~\cite{Brunken2009}. In  Table~\ref{tab-Qrot-HOCN} it can be observed that the relative differences between the two procedures are relatively small for temperatures above T~=~18.75~K. Nevertheless, it should be highlighted that, for temperatures from 37.5~K to 300~K, the direct sum results in slightly smaller values than the ones provided by the classical approximation. This fact could suggest that the direct sum has not been converged for this temperature interval. On the contrary, the direct sum has reached the convergence because, if this had not been reached at T~=~37.5~K, the difference between the direct sum and the classical approximation would have been increased for higher temperatures. In addition, we have compared the values obtained through the direct sum in this work and the rotational partition function values provided in the CDMS ~\citep{end2016}, which are also reported in Table~\ref{tab-PartF}. It can be observed that the values of the direct sum in this work and those from CDMS, which only provides values up to T~=~300~K, are practically the same. The relative difference between them is smaller than $0.00$~per~cent except for T~=~300~K, which is of $-0.04$~per~cent. Thus, it can be considered that the direct sum values of the rotational partition function have reached the convergence for all the temperature range up to T~=~300~K. However, at 500~K, according to the increasing difference between the direct sum and the classical approximation, we can state that the direct sum has not converged for this temperature. 
Therefore, even though the convergence of the direct sum seems to be reached up to T~=~300~K, we will consider for the rotational partition function in Eq.~(\ref{eq:rtvQapprox}) the values of the direct sum up to T~=~36~K inclusive and of the classical approximation from T~=~37~K up, according to the highest value.

The vibrational partition function has been obtained using the harmonic approximation from the  {\it ab initio} fundamental vibrational frequencies calculated  with the CCSD(T)/cc-pVQZ(all) level of theory~\citep{Mladenovic2008}. Their values are given in Table~\ref{tab-PartF} together with the values of rovibrational partition function computed with Eq.~(\ref{eq:rtvQapprox}) as well as  those up to 300~K of the CDMS rotational partition function. The relative difference between the results of this work and of CDMS  become larger than 10~per~cent at 225~K of temperature. Despite it is expected that the new values of the internal partition function are more precise, a more comprehensive list of experimental and predicted rovibrational data would be necessary to improve the partition function values as well as to get the convergence for the direct sum along all the interval up to T~=~500~K.

In the supplementary material, it is provided the rotational (as a direct sum up to T~=~36~K and from this temperature forward as a classical approximation), vibrational and rovibrational partition function computed in the present work for cyanic acid (HOCN) up to T~=~500 K using a 1 K interval.

\begin{table}
\caption{Rotational partition function for cyanic acid
  (HOCN). Comparison between the direct sum values and the classical
  approximation$^a$.}
\label{tab-Qrot-HOCN}
\begin{center}
{\footnotesize
\begin{tabular}{cccc}
\hline\hline
&   \multicolumn{3}{c}{HOCN}  \\ 
$T$(K) & $Q_{\rm rot}^{\rm approx}$$^b$ & $Q_{\rm rot}(\mbox{\rm direct sum})^c$ & Rel. Diff.(\%)$^d$ \\ \hline
\hline
    2.725 &     2.78 &    5.76 & 51.66 \\    
    5.000 &     6.92 &   10.30 & 32.83 \\         
    9.375 &    17.77 &   20.12 & 11.70 \\        
   18.750 &    50.26 &   50.82 &  1.09 \\         
   37.500 &   142.16 &  142.14 & -0.01 \\         
   75.000 &   402.08 &  401.38 & -0.18 \\         
  150.000 &  1137.25 & 1135.36 & -0.17 \\         
  225.000 &  2089.27 & 2087.44 & -0.09 \\         
  300.000 &  3216.64 & 3215.89 & -0.02 \\         
  500.000 &  6921.10 & 6863.28 & -0.84 \\   
\hline\hline
\end{tabular}
\begin{flushleft}
$^a$ Nuclear spin degeneracy is considered as $1$.\\
$^b$ Rotational partition function computed with the classical approximation using the rotational constants from \cite{Brunken2009}.

$^c$ Rotational partition function computed as a direct sum considering the predicted rotational energy levels up to
  $J$~=~72 and K$_a$~=~8~\citep{Brunken2009}. The predicted rotational energies are also given  in CDMS ~\citep{end2016}.\\

$^d$ Relative differences.

\end{flushleft}
}
\end{center}
\end{table}

\subsection{Fulminic acid (HCNO)}
\label{Q-HCNO}

Fulminic acid is a linear molecule of which experimental spectrum has been recorded in the millimeter-wave rotational and infrared ranges~\citep{Winnewisser1971,Albert1996,Schulze2000}. 
The internal partition function has been computed using Eq.~(\ref{eq:rtvQapprox}) where the rotational partition function has been calculated as a direct sum and compared in Table~\ref{tab-Qrot-HCNO} to the classical  
approximation for a linear molecule~\citep{Herzberg}.

The direct sum of the rotational partition function has been obtained considering the predicted rotational energies in the vibrational ground state up to $J$~=~90. The predictions have been calculated from the rotational Hamiltonian provided by \cite{Winnewisser1971} although, in this particular case, they have also been  reported in the CDMS ~\citep{end2016}.
The classical approximation has been calculated from the rotational constant $B$ provided by \cite{Winnewisser1971}. 
In Table~\ref{tab-Qrot-HCNO}, it can be observed that the classical approximation has a good agreement with the direct sum from T~=~18.75~K up and, although slightly smaller, the difference decreases for higher temperatures.

The harmonic approximation of the vibrational partition function has been obtained using the experimental fundamental vibrational energies~\citep{Albert1996,Schulze2000,NIST}. These values are given in Table~\ref{tab-PartF} together with the values of the internal partition function, written as product of the direct sum of the rotational contribution and harmonic approximation of the vibrational contribution, and the CDMS rotational partition function calculated as a direct sum~\citep{end2016}. In addition, the relative difference between the internal partition function obtained in this work and the CDMS partition function is also given. It can be noted that the relative diffference is larger than 20~per~cent at temperatures higher than 150~K. CDMS  does not provide the partition function at T~=~500~K.

In the supplementary material, it is reported the rotational (as a direct sum), vibrational and rovibrational partition function computed in the present work for fulminic acid (HCNO) up to T~=~500 K using a 1~K interval.

\begin{table}
\caption{Rotational partition function for fulminic acid
  (HCNO). Comparison between the direct sum values and the classical
  approximation$^a$.}
\label{tab-Qrot-HCNO}
\begin{center}
{\footnotesize
\begin{tabular}{cccc}
\hline\hline
&   \multicolumn{3}{c}{HCNO}  \\ 
$T$(K) & $Q_{\rm rot}^{\rm approx}$$^b$ & $Q_{\rm rot}(\mbox{\rm direct sum})^c$ & Rel. Diff.(\%)$^d$ \\ \hline
\hline
    2.725&     4.950696  &       5.298078  & 6.56 \\
    5.000&     9.083845  &       9.424730  & 3.62 \\
    9.375&    17.032210  &      17.369697  & 1.94 \\
   18.750&    34.064420  &      34.400567  & 0.98 \\
   37.500&    68.128841  &      68.466594  & 0.49 \\
   75.000&   136.257682  &     136.605301  & 0.25 \\
  150.000&   272.515363  &     272.904169  & 0.14 \\
  225.000&   408.773045  &     409.230843  & 0.11 \\
  300.000&   545.030726  &     545.585087  & 0.10 \\
  500.000&   908.384544  &     909.229250  & 0.09 \\
\hline\hline
\end{tabular}
\begin{flushleft}
$^a$ Nuclear spin degeneracy is considered as $1$.\\
$^b$ Rotational partition function computed with the classical approximation for linear molecules using the rotational constant from \cite{Winnewisser1971}.

$^c$ Rotational partition function computed as a direct sum considering the predicted rotational energy levels up to
  $J$~=~90~\citep{Winnewisser1971}. The predicted rotational energies are also  reported in the CDMS ~\citep{end2016}.\\

$^d$ Relative differences.

\end{flushleft}
}
\end{center}
\end{table}

\subsection{Isofulminic acid (HONC)}
\label{Q-HONC}

Isofulminic acid is the CHNO isomer with less spectroscopic information. This molecule is a non-linear asymmetric top molecule close to the prolate limit. 
Up to the present, the only experimental spectroscopic paper about this species provides some microwave data as well as {\it ab initio} calculations of the rotational parameters and the fundamental vibrational energies~\citep{Mladenovic2009}. 

Despite the lack of data, the rotational partition function has been calculated as a direct sum and this has been compared to the values given by the classical approximation (see Table~\ref{tab-Qrot-HONC}). The direct sum has been calculated
from the rotational energies up to $J$~=~80 and K$_a$~=~20 predicted with SPCAT code ~\citep{Pickett1991} using the experimental value of the rotational constant $B$ and the {\it ab initio} values of the rotational constants $A$ and $C$~\citep{Mladenovic2009}. These latter parameters have also been used for the classical approximation. The comparison between the direct sum and the classical approximation for the rotational partition function is also provided in Table~\ref{tab-Qrot-HONC}. From temperatures above 18.75~K, the relative difference is rather small and decreases for higher temperatures. Nevertheless, it can be noted that the value of the classical approximation becomes slightly larger (with a relative difference smaller than 0.1~per~cent) than the direct sum at T~=~500~K. This reveals that, at any temperature above, when this difference will become  substantially larger, the convergence of the direct sum obtained with the present spectroscopic data will not be reached anymore. 

The values of the harmonic approximation of the vibrational partition function are given
in Table~\ref{tab-PartF} together with the rovibrational ones calculated through Eq.~(\ref{eq:rtvQapprox}). The vibrational partition function is worked out using the fundamental vibrational frequencies determined with the {\it ab initio} CCSD(T)/cc-pCVTZ level of theory~\citep{Mladenovic2009}. To calculate the rovibrational partition function, the values of the rotational direct sum are considered up to T~=~437~K inclusive since the direct sum provides  larger values than those of the classical approximation. Above this temperature, the classical approximation is considered in its stead. In addition, in Table~\ref{tab-PartF} the internal partition function of this work is set side by side with the values of the rotational partition function reported in the CDMS  up to T~=~300~K~\citep{end2016}. It should be highlighted that the relative difference is larger than 10~per~cent at the temperature of 150~K. In any way, a more precise partition function could be determined in case that new experimental spectral measurements as well as experimental fundamentals are available.

In the supplementary material, it is reported the rotational (as a direct sum up to T~=~437~K and from this temperature  on as a classical approximation), vibrational and rovibrational partition function computed in the present work for isofulminic acid (HONC) up to T~=~500~K using a 1 K interval.

\begin{table}
\caption{Rotational partition function for isofulminic acid
  (HONC). Comparison between the direct sum values and the classical
  approximation$^a$.}
\label{tab-Qrot-HONC}
\begin{center}
{\footnotesize
\begin{tabular}{cccc}
\hline\hline
&   \multicolumn{3}{c}{HONC}  \\ 
$T$(K) & $Q_{\rm rot}^{\rm approx}$$^b$ & $Q_{\rm rot}(\mbox{\rm direct sum})^c$ & Rel. Diff.(\%)$^d$ \\ \hline
\hline
    2.725 &       2.782829&      5.552537 & 49.88 \\
    5.000 &       6.916585&      9.941600 & 30.43 \\
    9.375 &      17.757982&     19.740254 & 10.04 \\
   18.750 &      50.227159&     50.905466 &  1.33 \\ 
   37.500 &     142.063859&    142.724545 &  0.46 \\
   75.000 &     401.817274&    402.748482 &  0.23 \\
  150.000 &    1136.510876&   1137.826802 &  0.12 \\
  225.000 &    2087.903800&   2089.514569 &  0.08 \\
  300.000 &    3214.538189&   3216.363948 &  0.06 \\
  500.000 &    6916.584928&   6911.736014 & -0.07 \\
\hline\hline
\end{tabular}
\begin{flushleft}
$^a$ Nuclear spin degeneracy is considered as $1$.\\
$^b$ Rotational partition function computed with the classical approximation using the rotational constants from \cite{Mladenovic2009}.

$^c$ Rotational partition function computed as a direct sum considering the predicted rotational energy levels  up to
  $J$~=~80 calculated through SPCAT code~\citep{Pickett1991} using the rotational parameters from~\cite{Mladenovic2009}.\\

$^d$ Relative differences.

\end{flushleft}
}
\end{center}
\end{table}

\end{appendix}

\bsp	
\label{lastpage}
\end{document}